\documentclass[10pt]{article}

\usepackage[numbers]{natbib}
\usepackage{siunitx}
\usepackage{pdflscape}
\usepackage{float}
\usepackage{placeins}
\usepackage{bbm}
\usepackage{amsmath}

\usepackage{makecell}
\usepackage{algorithm2e}
\usepackage{authblk}
\usepackage{graphicx}
\usepackage{geometry}
\newgeometry{
textheight=9in,
textwidth=6.5in,
top=1in,
headheight=14pt,
headsep=25pt,
footskip=30pt,
left=.5in,
right=.5in
}
\usepackage[width=6.5in]{caption}

\begin{document}

\title{A two-step testing approach for comparing time-to-event data under non-proportional hazards}
\author[1,2]{Jonas Brugger}
\author[1]{Tim Friede}
\author[3]{Florian Klinglm\"{u}ller}
\author[2]{Martin Posch}
\author[2]{Robin Ristl}
\author[2,*]{Franz K\"{o}nig}
\date{February 13, 2024}

\affil[1]{\textit{\footnotesize University Medical Center G\"{o}ttingen, Department of Medical Statistics, G\"{o}ttingen, Germany}}
\affil[2]{\textit{\footnotesize Medical University of Vienna, Center for Medical Data Science, Institute of Medical Statistics, Vienna, Austria}}
\affil[3]{\textit{\footnotesize Austrian Agency for Health and Food Safety, Vienna, Austria}}
\affil[*]{\textit{\footnotesize Corresponding author, E-Mail: franz.koenig@meduniwien.ac.at}}

\maketitle

\abstract{The log-rank test and the Cox proportional hazards model are commonly used to compare time-to-event data in clinical trials, as they are most powerful under proportional hazards. But there is a loss of power if this assumption is violated, which is the case for some new oncology drugs like immunotherapies. We consider a two-stage test procedure, in which the weighting of the log-rank test statistic depends on a pre-test of the proportional hazards assumption. I.e., depending on the pre-test either the log-rank or an alternative test is used to compare the survival probabilities. We show that if naively implemented this can lead to a substantial inflation of the type-I error rate. To address this, we embed the two-stage test in a permutation test framework to keep the nominal level alpha. We compare the operating characteristics of the two-stage test with the log-rank test and other tests by clinical trial simulations.}

\section{Introduction}

Time-to-event type endpoints are commonly used in confirmatory clinical oncology trials. Examples include the traditionally used overall survival (OS) or surrogate outcomes like progression-free survival (PFS) or disease-free survival (DFS) that might also reflect on the quality of life or save resources due to higher event rates and consequently, shorter observation periods and overall trial duration. In survival analysis, essential tools to compare survival probabilities or hazard functions in a two-arm randomized controlled setting include the log-rank test \cite{mantel1966evaluation} \cite{peto1972asymptotically} and Cox proportional hazards model \cite{cox1972regression}. They rely on the assumption that the treatment's effect is time-invariant. However, if the assumption of proportional hazards (PH) is violated, they both might lack power and the interpretation of their results becomes ambiguous. This suggests a requirement for methods better suited to handle such scenarios. There is an ongoing debate about the application of appropriate methods under non-proportional hazards (NPH) and as of today there is no consensus about best practices. Examples for detailed reviews on this subject include \cite{ananthakrishnan2021critical} and \cite{bardo2023methods}. Performance of popular methods has been investigated in several simulation studies such as \cite{callegaro2017testing, shen2023nonproportional, klinglmuller2023neutral}.

\begin{figure}[!ht]
\centering
\includegraphics[width=.9\textwidth]
{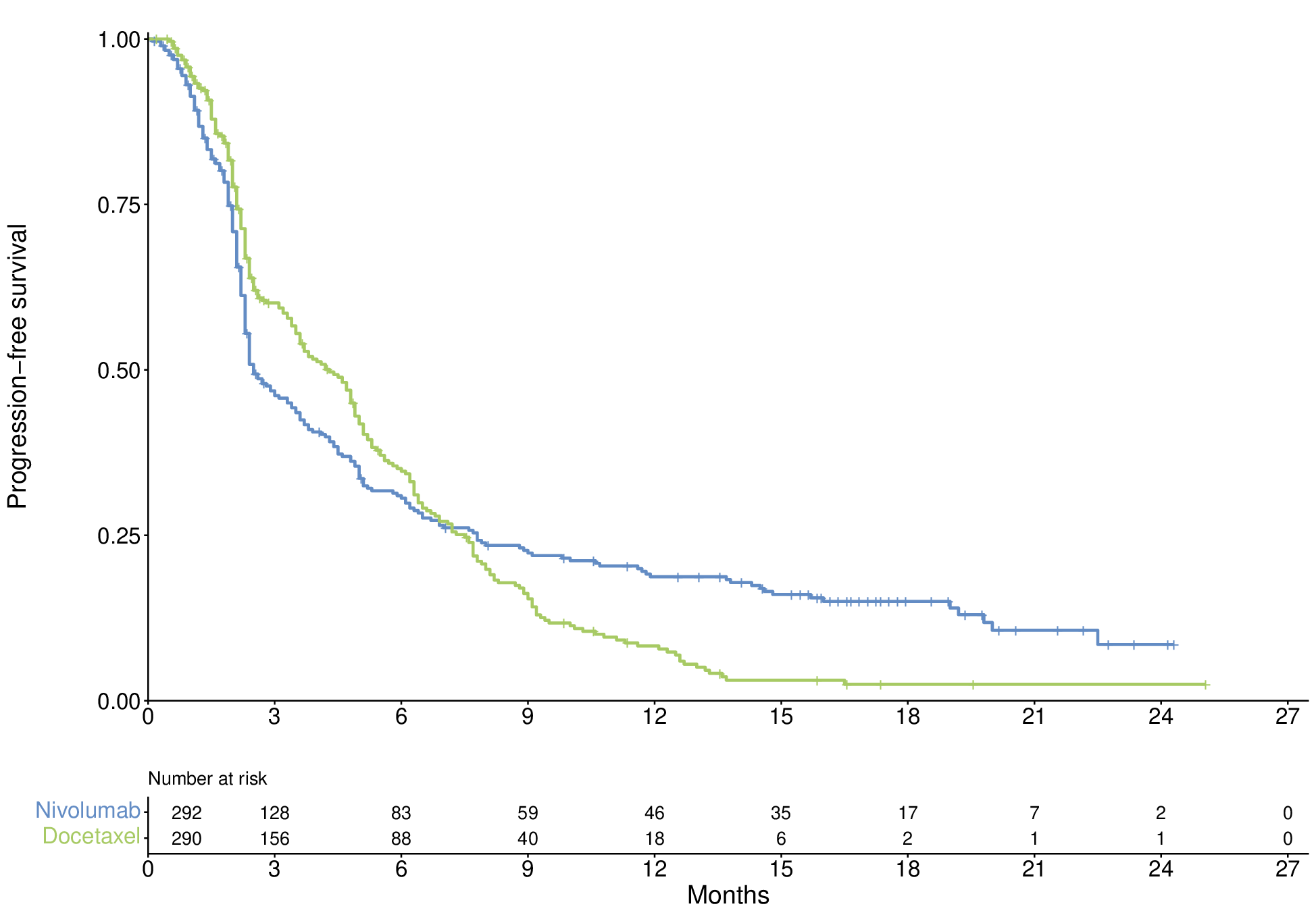}
\caption{\textbf{Case study:} Reconstructed Kaplan-Meier estimates of progression-free survival from a trial of Nivolumab in patients with non-small cell lung cancer.  \cite{borghaei2015nivolumab}. We digitized the KM-curves using the webplotdigitizer \cite{rohatgi2022} and estimated the original data using Guyot´s method \cite{guyot2012enhanced}. In the original study, a stratified log-rank test gave a two-sided p-value of 0.39 for the comparison of PFS.}
\label{nivolumab}
\end{figure}

The unique mechanisms associated with novel oncology treatments have generated a growing demand for statistical methods designed to address situations involving non-proportional hazards. One major catalyst for this discussion revolves around immunotherapeutic agents, which enhance the immune system's ability to target cancer cells. They are renowned for exhibiting response delays, meaning potential benefits become noticeable only some time after treatment initiation. We selected a motivating example of this effect from a clinical trial in which patients with advanced melanoma received ipilimumab, a monoclonal antibody targeting receptors that suppress the immune response \cite{hodi2010improved}. The authors found that overall survival of the cohort treated with ipilimumab was significantly extended, however the Kaplan-Meier curves suggest that survival probabilities are roughly equivalent for a period following after administration. In a trial assessing the effectiveness of nivolumab versus docetaxel in patients with non-small cell lung cancer, the initial estimated survival probability was even inferior. However, nivolumab was favourable in terms of long-term survival. The same was true for the secondary endpoint of PFS \cite{borghaei2015nivolumab}, however this result was non-significant in the original analysis. To provide an illustrative example, we reconstructed the original Kaplan-Meier estimates for PFS in Figure \ref{nivolumab} and analyzed it as a case study using the methodology from this manuscript.

Without specification of the underlying mechanism we investigate three common patterns of NPH in this work. In addition to the aforementioned delayed effect, we study subgroup effects of subjects in the treatment cohort, e.g., based on an (unknown) biomarker. One such example was observed in a study of gefitinib, an inhibitor of epidermal growth factor receptor (EGFR) signaling, versus carboplatin in patients with pulmonary adenocarcinoma. Gefitinib proved effective against tumors with a EGFR gene mutation that leads to their over-expression and was inferior to carboplatin in the remaining cohort with regards to the primary endpoint of PFS. Overall, this led to a crossing of survival curves \cite{mok2009gefitinib}. Lastly, we consider a scenario in which random disease progression events like metastasizing of a tumor can increase the hazard rate and negate potential treatment benefits in a proportion of patients.

In instances where the PH assumption is violated, modifications of the log-rank test have been shown to increase its power. The idea is to focus on the points in time where the treatment effect is expected to be most prominent and weighting observations accordingly. A well-known class of functions, introduced by Fleming-Harrington, allows for focusing on specific intervals within the observation period \cite{harrington1982class}. Similarly, modestly weighted tests have proven advantageous in terms of power in instances of late treatment effects \cite{magirr2019modestly}. When there is a clear understanding of both the timing and extent of the treatment's benefit, it is recommended to specify an appropriate analysis method in the protocol based on that knowledge. In alternative scenarios, maximum type test statistics can serve as a robust alternative to using individual weighting functions or the log-rank test alone \cite{lin2020alternative}.

In this study, we explore a log-rank-based method designed to offer flexibility  similar to combination tests. In a two-step approach, we condition the second-step test for comparing survival probabilities on an evaluation of the PH assumption via the Grambsch-Therneau (GT) test \cite{grambsch1994proportional}. If it fails to reject, we proceed with a standard log-rank test, otherwise we opt for a pre-defined alternative test, e.g. a weighted test instead. We assess performance measures with regards to the pre-test significance level and consider a permutation test adjustment to account for potential type-I error inflation of the considered two-stage testing procedure. A similar framework was previously explored by Campbell and Dean \cite{campbell2014consequences}, who considered Cox regression-based tests, a conditional log-rank test and accelerated failure time models as second-step tests under scenarios of non-proportional hazards.

The remainder of this manuscript is structured as follows: We shortly discuss the log-rank test as well as weighted- and maximum type alternative that are more suitable under scenarios of NPH in the methods section. After that, we introduce the concept of the two-step test and a permutation test framework to manage its type-I error inflation. Finally, we investigate the performance measures of this approach with diverse alternative tests under NPH in a simulation study and by means of a reconstructed real-life example and discuss the results.

\section{Methods}

\subsection{(Weighted) log-rank tests}

We consider the weighted log-rank test statistic \cite{fine2007consequences}

\begin{align*}
    Z = \frac{\sum_{t \in j} w_t [r_{t1} - r_{t} \frac{n_{t1}}{n_t}]}{\sqrt{\sum_{t \in j} w_t^2 r_{t} \frac{n_{t1}}{n_t}(1 - \frac{n_{t1}}{n_t}) \frac{n_{t} - r_{t}}{n_t - 1}}}
\end{align*}

where $j \in \{1, \ldots J\}$ denotes the set of observed distinct event times. $n_{j1}$ and $r_{j1}$ are the number at risk and the the number of observed events in the experimental group at time $j$ and $n_{j}$ and $r_{j}$ their total amount in both groups. The standard log-rank test is the special case where the weights are $w_t = 1$ for all $t$ \cite{mantel1966evaluation} \cite{peto1972asymptotically}. Under proportional hazards, the log-rank test has been shown to be the most powerful rank-invariant test for the usual null-hypothesis that the hazard functions of control and treatment group are identical \cite{peto1972asymptotically}. In other scenarios, the power can be maintained by appropriately weighting observations. Two common weighting functions include the Fleming-Harrington (FH) $G^{\rho, \gamma}$ class and the class of modest weights by Magirr and Burmann. FH weighting functions are defined as \cite{harrington1982class}

\begin{align*}
    w(j, \rho, \gamma) := \hat{S}(j^-)^\rho(1 - \hat{S}(j^-))^\gamma
\end{align*}

where $\hat{S}(j^-)$ denotes the pooled KM-estimator of survival probability right before $j$. In what intervals after the time from randomization observations are emphasized depends on the choice of $\rho$ and $\gamma$. In the following sections, we use $FH (\rho, \gamma)$ as an abbreviation for weighted tests of the Fleming-Harrington class with the associated hyperparameters. Arbitrary weighting functions in general and those of the $G^{\rho, \gamma}$ that concentrate on late stages of the observation period have been shown to be type-I error inflated in a one-sided test setting if the hazard of the treatment group is only favourable in these segments (see for example \cite{magirr2021non}). It is possible for hazard functions to cross, but the control group to have strictly favourable survival probability. Modest weighting functions emphasise late events while controlling the type-I error probability under the scenario of crossing hazard functions. Their weights \cite{magirr2019modestly}

\begin{align*}
   w(j, t^*):= \frac{1}{max\{\hat{S}(j), \hat{S}(t^*)\}} \\
\end{align*}

are calculated similarly as a function of the pooled KM-estimator and a hyperparamter $t^*$, which controls the degree by which late events are highlighted over ones at the beginning of the observation period. The larger $t^*$ the longer the weights increase. After $t^*$ the weights are constant.
Modestly weighted log-rank tests were developed for situations involving delayed treatment effects, where they have demonstrated high power. However, in general performance of weighted tests are highly scenario specific \cite{royston2020simulation, klinglmuller2023neutral}.

\subsection{Maximum-type tests}

In situations where occurrence and intensity of non-proportional hazards is difficult to predict, tests incorporating different weight schemes simultaneously may offer a robust alternative. One well known example is the maximum combination (max-combo) test, which identifies the largest test statistic associated with $k$ different weighting functions \cite{lee1996some}. An adjusted $p$-value is typically calculated from the multivariate distribution of the respective test statistics. Within combination tests, weight functions are usually deliberately selected to account for different patterns of potential treatment effects. The max-combo test is commonly applied in conjunction with weighting functions of the FH class. Lee \cite{lee2007versatility} proposed using $FH (1, 0)$ and $FH (0, 1)$ to consider both differences in early and late survival probability. Lin et al. \cite{lin2020alternative} suggested a combination of the four weighting functions $FH (0, 0)$, $FH (1, 0)$, $FH (0, 1)$ and $FH (1, 1)$, that includes the standard log-rank test. We used the one proposed by Lin throughout our simulation study if not specified otherwise.

\section{Proposed two-step test} \label{twostep_section}

Here we consider a framework in which the test of a treatment's effect is based on prior information on whether the proportional hazards assumption is reasonable. The Grambsch-Therneau (GT) test \cite{grambsch1994proportional}, a generalization of Schoenfeld´s approach \cite{schoenfeld1982partial}, does this by evaluating trends of the scaled Schoenfeld residuals in Cox regression model. The null-hypothesis of proportional hazards may or may not be rejected depending on whether or not the p-value is smaller than some significance level, which we term $\alpha_{pre}$ to differentiate between the significance level of the actual comparison of the TTE-endpoint. Given $\alpha_{pre}$, the second-step test under PH is always the standard log-rank test in this framework. In case the GT-test rejects, an alternative pre-defined test is performed, e.g., a weighted test or combination of weighted tests.

\begin{figure}[!ht]
\centering
\includegraphics[width=13cm]{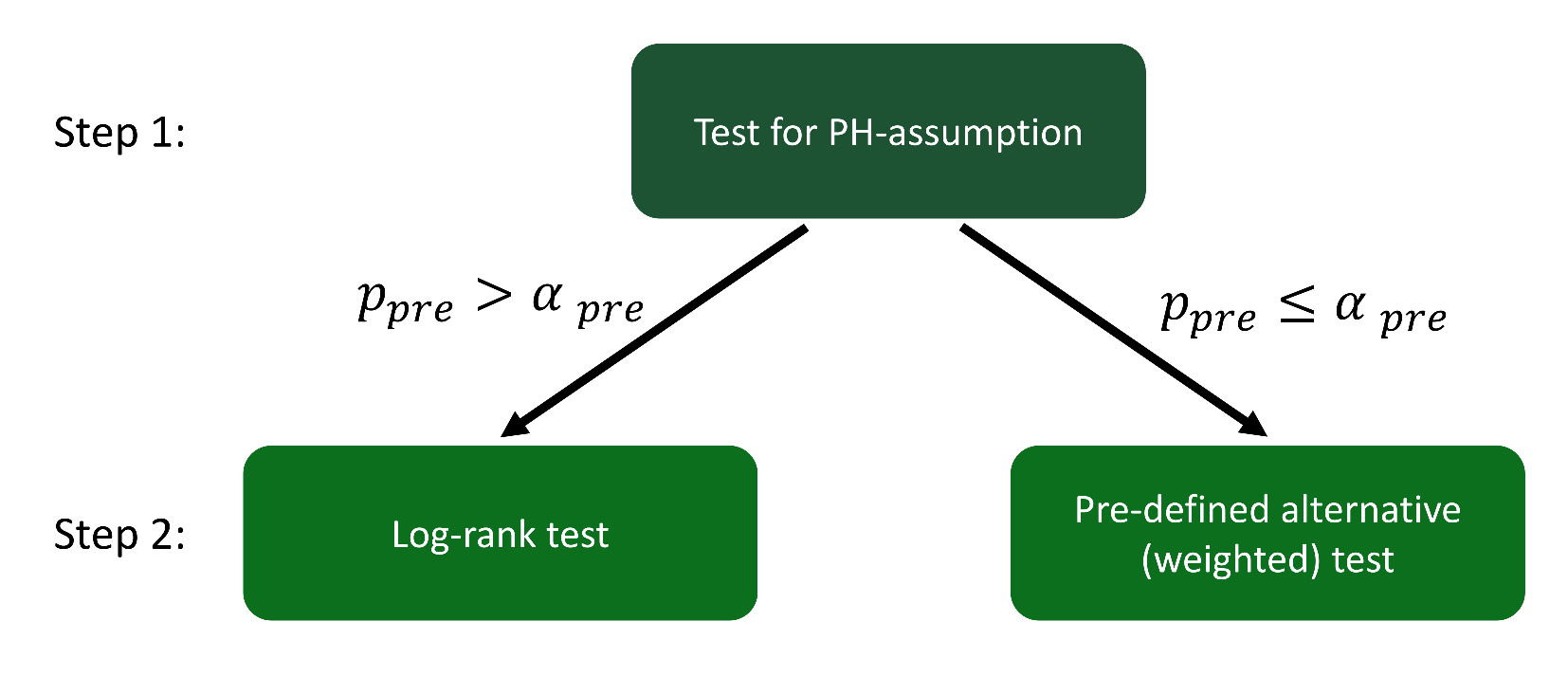}
\caption{\textbf{Two-step test principle: A visual overview.} In a first step the proportional hazards assumption is evaluated. In a second step the subsequent test of group-specific survival probabilities is conditioned on the rejection or non-rejection of the PH-assumption at significance level $\alpha_{pre}$.}
\label{twostep}
\end{figure}

In the following sections, this simple framework will be referred to as the naive two-step (nTS) test. A visualization of the procedure is provided in Figure \ref{twostep}.

It is well-known that without adjustment of the significance level, sequential test procedures like this two-step approach generally do not maintain the intended type-I error probability. The exact divergence depends on the correlation of the two test statistics \cite{royston2014approach}. One objective of our simulation study was to show how type-I error rate is related to the significance level of the pre-test $\alpha_{pre}$. It has to be noted, that the higher $\alpha_{pre}$, the more frequently we will switch from the log-rank test to the alternative second-step test. In addition to the naive approach, we implement a permutation test adjustment of the two-step test that is capable of controlling type-I error rates under the null-hypothesis.

\subsection{Permutation test adjustment of the two-step test}

We address a potential type-I error inflation by implementing a permutation test to obtain unbiased estimates of the specific quantiles of the null-distribution by Monte Carlo sampling (see for example \cite{edgington2007randomization}). In this work, we consider a top-down approach, in which the entire two-step test procedure is applied on data with re-sampled treatment labels. Performing the permutation two-step (pTS) test requires the following steps:

\begin{enumerate}
    \item Define the significance levels $\alpha_{pre}$ and $\alpha$  for the pre-test (Step 1 in Figure \ref{twostep}) and final test (Step 2 in Figure \ref{twostep}), respectively.
    \item Calculate $p_0$, the p-value of the nTS test for comparing the time-to-event data of the two groups (Step 2 in Figure \ref{twostep}). Depending on the outcome of comparing the pre-test p-value $p_{pre}$ to $\alpha_{pre}$ either the p-value of the log-rank or the alternative test will be taken for $p_0$.
    \item Randomly permute the treatment covariate $m$ times and calculate the two-step tests p-value for each of the post-hoc sequences. We denote the resulting p-values as $p_1,\ldots, p_{m}$.
    \item Calculate the overall pTS test's p-value as $p = \frac{\sum_{i=1}^{m} \mathbbm{1}_{p_i < p_0}}{m}$
    \item Reject the null-hypothesis if $p \leq \alpha$. 
\end{enumerate}

We included an algorithm of the procedure in appendix \ref{alg:random}. In the following simulation study, we investigate the type-I error of naive and permutation test under the null-hypothesis and compare their power to the standard log-rank test and modifications in selected scenarios.

\section{Simulation Study}

\subsection{Simulation setup}

We looked at three different sources of non-proportional hazards in addition to proportional hazards scenarios: Delayed treatment effects, subgroup effects and random disease progression events that can lower the hazard rate of subjects in both trial arms. Throughout the entire simulation study we assume a randomized controlled trial of equally sized groups. With regards to study design we differentiate between design choices and assumptions, analogous to \cite{meyer2022decision}. Design choices are characteristics determined by the researchers like the number of total subjects and the number of events after which the trial terminates. Subject number varies between 400 and 700 in our simulations, with 400 being the reference scenario. The required number of events for trial termination is calibrated such that the log-rank test has around 80\% power in a reference scenario given that the alternative hypothesis is true. We further assume that the length of the observation period is event-driven, meaning that trials do not terminate after a given amount of time. On the other hand, factors that are not (entirely) in control of the researchers like the recruitment speed or the censoring pattern, are summarized as assumptions. We investigated three different recruitment speeds summarized as slow, medium and fast recruitment for convenience and defined the medium speed as the reference case. We assume that no censoring occurs besides administrative censoring when the pre-planned number of events is reached. Implicitly, the results of our simulation study are only valid in instances where censoring is non-informative. Median survival of the control group is fixed at 12 months, the treatment effect size is scenario specific. The reference scenarios are defined as follows:

\begin{itemize}
\item Proportional hazards scenario: The median survival of the treatment group is fixed at 18 months.
    \item Delayed treatment effect: In this scenario the hazard rate of the treatment group is identical to the one of the controls initially. After the effect onset, the conditional median survival of the experimental group is 18 months provided the subject survived the delay period. We further distinguish between a \textbf{short effect delay} of 2 months and a \textbf{long effect delay} of 4 months.
    \item Disease progression scenario: Random disease progression events can lower the hazard rates of subjects in both trial arms. Progression events follow an exponential distribution with a median time of 36 months in both study groups. This implies that around 20\% of patients experience a progression event within 1 year of inclusion. We consider two different scenarios based on the conditional hazards after disease progression. In the scenario of \textbf{long survival after progression} the conditional median survival is 9 months, in case of \textbf{short survival after progression} the conditional median survival time is 3 months. Without disease progression events, the median survival time in the experimental group is 18 months.
    \item Subgroup scenario: Here, a proportion of subjects in the treatment group are assumed to respond differently to the drug. We therefore express the treatment effect as a function of two hazard functions or hazard ratios with regards to the control group: the treatment subgroup's and their complement's. In our example, the treatment subgroup is assumed to have a substantially larger survival probability with a median survival time of 120 months. The complement's median survival time of 12 months is the same as the control's. In the discussion of results, hazard ratios are with regards to the complement of the treatment subgroup compared to the control if not specified otherwise. Depending on the percentage of patients in the subgroup we again distinguish between \textbf{low subgroup prevalence} of 20\% and a \textbf{high subgroup prevalence} of 50\%.
    \item Null-scenario: The experimental and control group have equivalent median survival times of 12 months. The study terminates after $\frac{3}{4}$ of the participants experienced an event.

\end{itemize}

Here, we should note that these values only correspond to the reference scenario. A complete list of alternative study characteristics we considered can be found in table \ref{tabsimparams}. In addition, we varied the hazard rate of the treatment group after effect onset, without disease progression events or in the subgroup-complement (depending on the scenario) within the simulations under the alternative hypothesis. For a look at the exact survival distributions under the reference scenario we refer to the appendix (Figure \ref{scenarios}). Throughout this simulation study, we only consider one-sided tests. Under the assumption of proportional hazards, the one-sided null- and alternative hypothesis are defined as follows:

\begin{align*}
    H_0: h_0(t) &\leq h_1(t) \text{ for all } t \\
    H_1: h_0(t)  &> h_1(t) \text{ for all } t
\end{align*}

where $h_0(t)$ and $h_1(t)$ denote the hazard functions of the control and treatment group, respectively. Note that even in case of a treatment benefit, the inequality $h_0(t) > h_1(t)$ might not be true for all time points under NPH. Nevertheless, superior treatment effect can generally be recognized by a predominately favourable hazard that translates into better survival probability. The one-sided significance level is kept at $\alpha = 0.025$ throughout.

In each scenario, the standard log-rank test serves as the reference method for power comparison. In addition, we analyzed simulated data using the following methods:

\begin{table} 
    \centering
    \begin{tabular}{|p{1.3in}|p{2.1in}|p{2.6in}|} \hline 
    \makecell{\textbf{Name}} & \makecell{\textbf{Investigated values}} & \makecell{\textbf{Description}}\\
    \hline
    Number of subjects & \textbf{400}, 700 & Number of subjects to be included in the trial before recruitment stop. \\ 
    Number of events & Calibrated for the log-rank test to have approximately 80\% power under the reference scenario & Number of events that mark the end of the observation period. \\
    $\alpha_{pre}$ & 0-1 in steps of 0.025 for scenarios under the null-hypothesis, \newline \{0.05, \textbf{0.2}\} under scenarios of the alternative hypothesis  & Significance level of the pre-test for the proportional hazards assumption. This significance level is always two-sided. \\
    Recruitment stop & 700, \textbf{400} and 100 days, in the following slow, \textbf{medium} and fast recruitment & Length of recruitment period after trial start. The recruitment is assumed to be uniform in the respective time interval, therefore the speed is inversely proportional to the length of the window.  \\
    \hline

\end{tabular}
    \caption{\textbf{Design choices and trial assumptions that are varied within the simulation study.} The values of the reference scenario used for the calibration of required events and visualization are stated in bold. Only results for these scenarios are presented in the main part of the manuscript.}
    \label{tabsimparams}
\end{table}

\begin{itemize}
    \item \textbf{FH (1, 0):} Weighted log-rank test of the Fleming Harrington class with $\rho = 1$ and $\gamma = 0$. In later sections, this method will also be referred to as the weighted log-rank test with early emphasis.
    \item \textbf{FH (1, 1):} Weighted log-rank test of the Fleming Harrington class with $\rho = 1$ and $\gamma = 1$. In later sections, this method will also be referred to as the weighted log-rank test with mid-emphasis
    \item \textbf{FH (0, 1):} Weighted log-rank test of the Fleming Harrington class with $\rho = 0$ and $\gamma = 1$. In later sections, this method will also be referred to as the weighted log-rank test with late emphasis.
    \item \textbf{Max-combo test} as proposed in Lin et al \cite{lin2020alternative}.
    \item \textbf{Modest (6):} Modestly weighted log-rank test with hyperparameter $t^{*} = 6$ months. 
    \item \textbf{Modest (12):} Modestly weighted log-rank test with hyperparameter $t^{*} = 12$ months. 
\end{itemize}

Moreover, we performed two-step tests with the above methods as the alternative in case of rejection of the PH-assumption. In the two-step framework, the max-combo test in the second stage does not include the standard log-rank test as it is the alternative method under proportional hazards. For scenarios of the alternative hypothesis, we looked into pre-test levels of $\alpha_{pre} \in \{0.05, 0.2\}$. The rationale behind this was to incorporate a higher significance level that permits a more lenient rejection of the null hypothesis of PH, alongside the conventional level. Under scenarios of the null-hypothesis, the rejection probabilities of the standard two-step tests are estimated as a function of the pre-test significance level.

We applied the KM-transform of the time axis when testing the proportional hazards assumption via the GT-test, since this does not impose a functional form of the hazard ratio function and is robust to outliers \cite{therneau1997extending}.
We evaluated power of the two-step tests as a function of the hazard ratio of the treatment group or the subgroup complement in case of the subgroup scenario.

The number of replications was chosen with the intent of obtaining a sufficiently small Monte Carlo error of the rejection probability. When analyzing scenarios under the alternative hypothesis, we simulated 10000 replications, which guarantees a maximum standard error of $SE = \sqrt{\frac{0.5^2}{10000}} = 0.005$ in the event of the worst case scenario of 50\% rejection probability. Therefore, the standard error of the estimator of the rejection probability is at most 0.5 percentage points in any scenario. We performed 100000 replications for the null-hypothesis scenario, meaning the standard error of the type-I error rate is $\approx 5\cdot 10^{-4}$ at a one-sided significance level of $\alpha = 0.025$. Within each application of the pTS test we performed $m = 2500$ permutations on realized data sets. We decided on this number after reviewing similar applications of permutation tests \cite{campbell2014consequences} \cite{royston2016augmenting}.

All calculations were performed using R, version 4.2.2 or higher \cite{Rprogram}. We simulated data using the packages \textit{SimNPH} \cite{klinglmuller2023neutral} and \textit{SimDesign} \cite{SimDesign}. We calculated weighted log-rank tests with the function \textit{wlrt} from the \textit{nphCRT}-package \cite{nphRCT}, and max-combo tests using \textit{logrank.maxtest} from the \textit{nph}-package \cite{ristl2021delayed}.

\subsection{Simulation results}

In the subsequent section, we present findings related to the null-hypothesis scenario, proportional hazards, and treatment subgroups. Furthermore, we limited our focus in the main section to the outcomes of $\alpha_{pre} = 0.2$. For the results of the delayed effect and disease progression scenarios, as well as a lower pre-test significance level we refer to the appendix and online supplement. Under the alternative scenario, the estimated power is shown as a function of the hazard ratio of the treatment group or subgroup complement in case of the subgroups scenario compared to the control group. A $HR < 1$ implies a favourable effect of the experimental treatment. We simulated scenarios of higher sample size and different recruitment speed only for the naive tests due to the computational complexity of the pTS tests. These results can be found in the supplementary material. We assume that performance differences of the nTS tests under different sample size and recruitment speed apply to the pTS tests in a similar way.

\subsubsection{Power of the naive two-step (nTS) test}

\begin{figure}[!ht]
\centering
\includegraphics[width=.85\textwidth]{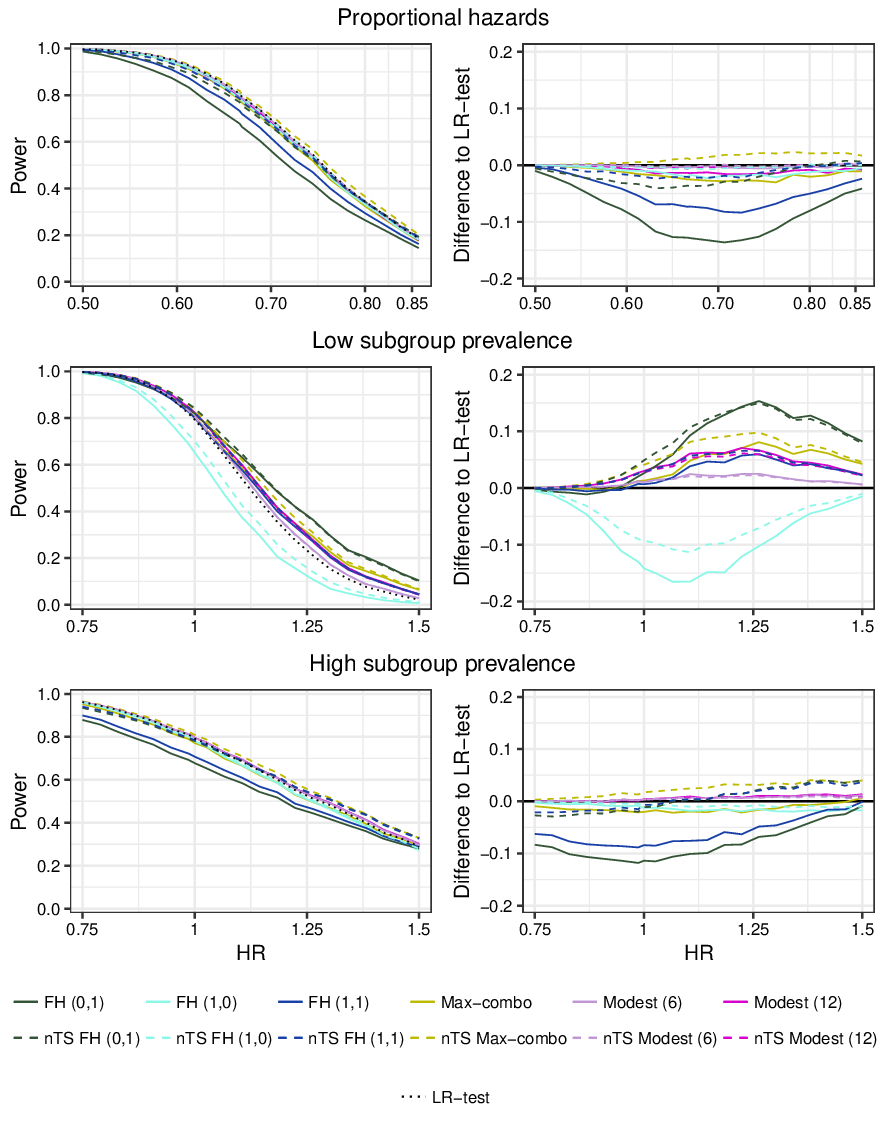}
\caption{\textbf{Proportional hazards and subgroup scenarios:} Absolute power (left column) and power difference (right column) to the log-rank test of the naive two-step ("nTS") tests. Results of the two-step tests are displayed as dashed lines in the same colour as the associated conventional approaches under NPH. The power of the log-rank test ("LR-test") serves as a reference and is displayed as a black dotted line. In the subgroup scenarios, the hazard ratio (HR) is with regard to the complement of the treatment subgroup. The HR of the treatment subgroup compared to the control is always 0.1 (corresponding to a median survival time of 120 months).}
\label{res:main}
\end{figure}

Statistical power as a function of the hazard ratio in the aforementioned scenarios is illustrated in Figure \ref{res:main}. The log-rank test is the most powerful of the conventional tests under PH over the majority of the domain. Only the modestly weighted tests had similar rejection rates in some instances. The nTS max-combo test, however, manages to improve on the power of the log-rank test particularly under scenarios of smaller treatment effects. Other two-step tests also have increased power in this area.

In the subgroup scenarios, the two-step tests were frequently superior to both their associated conventional second-step tests, but the benefit heavily depended on the effect size in the general treatment group. In the instance of a small prevalence, the nTS FH (0, 1) substantially improved on the power of the two associated second-step tests if the survival probability of the experimental group is strictly better. The same is true for the nTS max-combo test, only its power is higher in all parts of the domain. Aside from those, the power of the remaining two-step approaches is mostly a compromise between the ones of the associated conventional tests.

\FloatBarrier

\FloatBarrier

Under the scenario of high subgroup prevalence, we observed similar results. Again, many of the two-step tests generally improve on the power of both the associated conventional tests. For the nTS max-combo test this is again strictly the case. If the hazard functions cross, the power gain of the two-step tests over the conventional methods is most noticeable.

\subsubsection{Type-I error of the naive two-step (nTS) test}

\FloatBarrier

\begin{figure}[!ht]
\centering
\includegraphics[width=.9\textwidth]{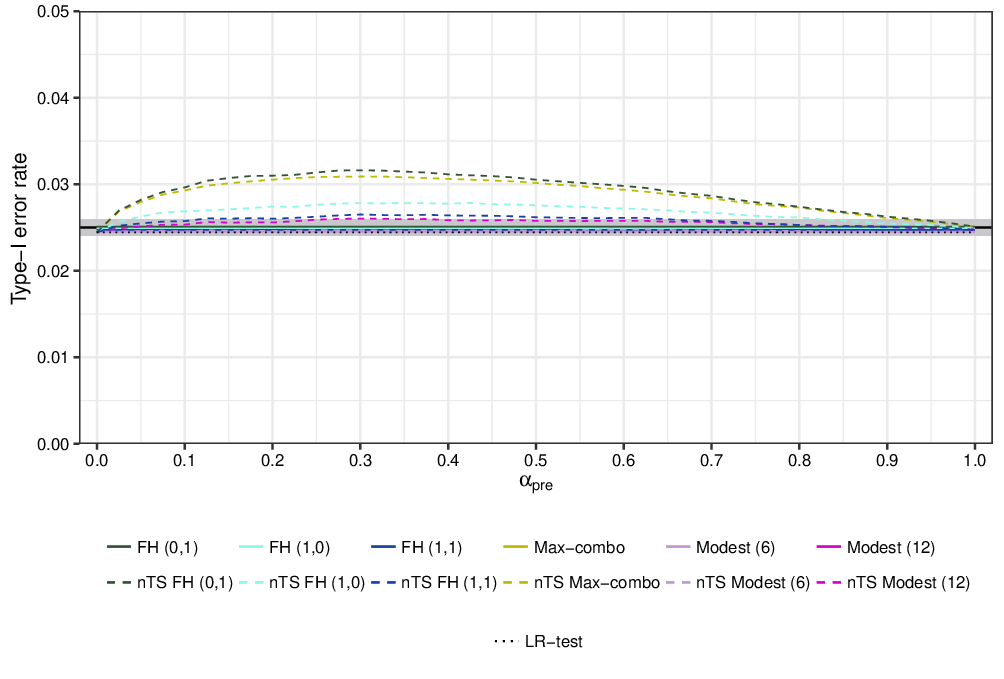}
\caption{\textbf{Null-hypothesis:} Type-I error rates as function of the pre-test level. The 95\% prediction interval of the estimated type-I error rate under identical hazard functions is marked in grey. Values outside this area provide strong evidence of type-I error inflation. The one-sided significance level $\alpha = 0.025$ is indicated as a solid black line. The naive two-step  test procedures ("nTS") are represented by dashed lines in the same colour as the associated conventional approaches. The results of the log-rank test ("LR-Test") are shown as a black dotted line. }
\label{res:nullidentical}
\end{figure}

\FloatBarrier

\begin{figure}[!ht]
\centering
\includegraphics[width=.9\textwidth]{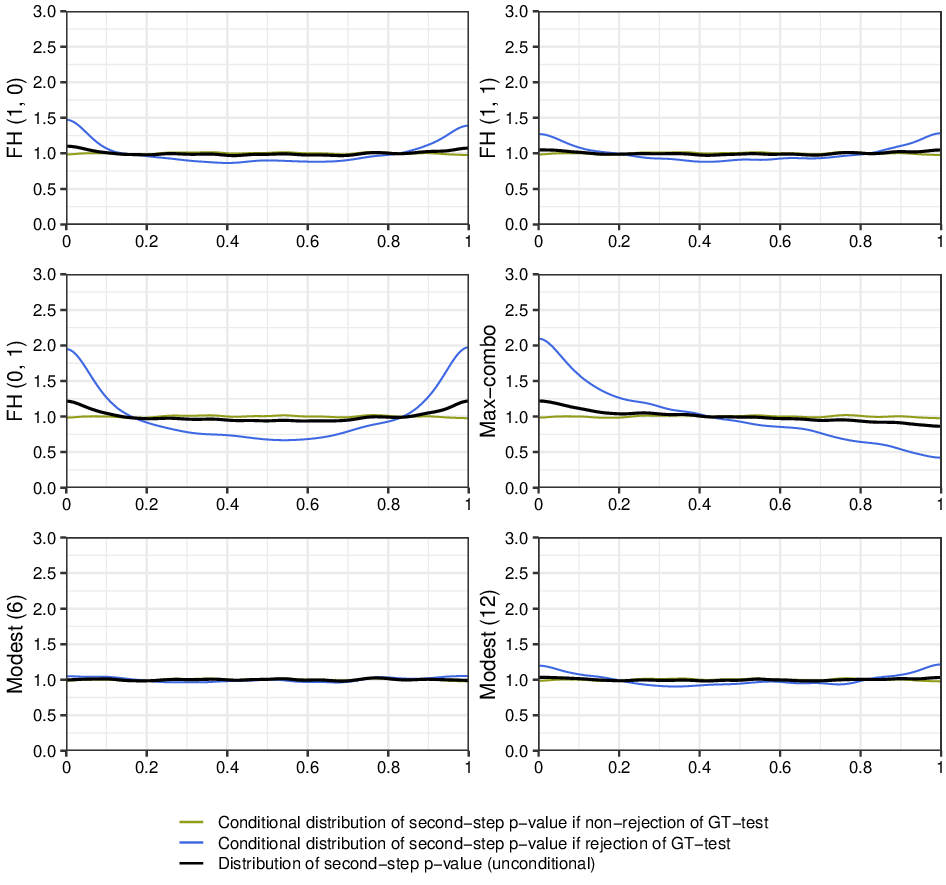}
\caption{\textbf{Null hypothesis:} Kernel density estimation of the naive two step test's p-values conditioned on the results of the pre-test for PH. The pre-test significance level is $\alpha_{pre}=0.2$. The y-axis titles denotes the second-step test under NPH. The second-step test in case of non-rejection of the PH assumption is always the log-rank test.}
\label{res:corplot}
\end{figure}

Type-I error rate of the nTS tests with regards to the pre-test significance level under the null-scenario of identical survival curves is displayed in Figure \ref{res:nullidentical}. Note that the rate is constant for the conventional methods as they are independent of the pre-test. Assuming $\alpha_{pre} = 0$, the PH-assumption is never rejected by the pre-test, therefore the type-I error rates of the nTS tests are identical to the corresponding conventional tests. However, in case of a different pre-test level the nTS tests generally do not maintain type-I error control. Our simulations provide strong evidence that this is the case for at least some pre-test significance levels for all nTS tests except the two-step Modest (6). The nTS FH (0, 1) and nTS Max-combo test are the most inflated with rejection probabilities of $\geq 0.03$ for $\alpha_{pre} \in (0.15, 0.5)$
Figure \ref{res:corplot} displays the conditional distribution from kernel density estimators of the second-step tests p-value given the pre-test's decision. In this figure, the pre-test significance level is fixed at $\alpha_{pre} = 0.2$. Under the assumption that tests of different steps in the procedure are independent, we would expect the distributions to be uniform. There is no clear pattern violation to be found in the distribution of the standard log-rank test's p-value in black. However, when the GT-test suggests NPH, the p-values of the weighted test have increased mass at the extremes, indicating a type-I error inflation. The direction of the correlation of the two test statistics likely depends on whether or not a weighting function is "appropriate" for a given instance. It is reasonable that it is strictly positive in case of the Max-combo test, which is robust for various hazard functions. The unconditional distribution can be thought of as a weighted sum as the GT-tests p-value is uniform under the null.

\FloatBarrier

\subsubsection{Performance measures of the permutation two-step (pTS) test}

We included in the supplementary material a visualization of the type-I error rates of the permutation tests as a function of $\alpha_{pre}$, analogous to Figure \ref{res:nullidentical}. At no point does the type-I error rate lie outside the 95\% confidence region. We therefore conclude that they are able to maintain type-I error control in the scenario of identical survival functions.

\begin{figure}[!ht]
\centering
\includegraphics[width=.85\textwidth]{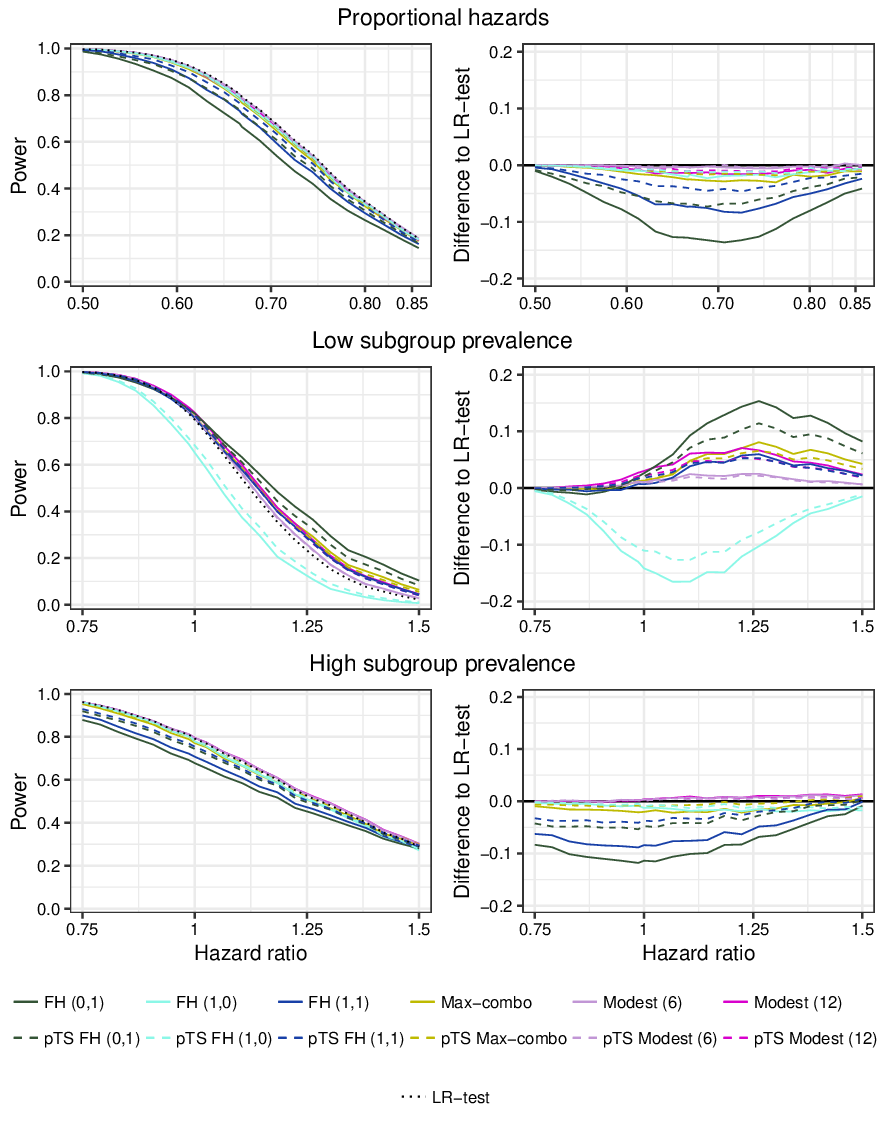}
\caption{\textbf{Proportional hazards and subgroup scenarios:} Absolute power (left column) and power difference (right column) to the log-rank test of the permutation two-step ("pTS") tests. Results of the two-step tests are displayed as dashed lines in the same colour as the associated conventional approaches under NPH. The power of the log-rank test ("LR-test") serves as a reference and is displayed as a black dotted line. In the subgroup scenarios, the hazard ratio (HR) is with regard to the complement of the treatment subgroup. The HR of the treatment subgroup compared to the control is always 0.1 (corresponding to a median survival time of 120 months).}
\label{res:randommain}
\end{figure}

The power of the pTS tests in the proportional hazards and subgroup scenarios compared to the conventional tests is illustrated in Figure \ref{res:randommain}. It is evident, that they lose some power compared to the naive approaches irrespective of the scenario. As expected, the standard log-rank test is almost strictly the most powerful test under PH. Rejection rates of the pTS tests almost consistently lie between the log-rank test and the corresponding conventional test under NPH.
For the most part, this is also the case in the subgroup scenarios. The pTS tests manage to outperform the conventional methods only in selected parts of the domain. In the low subgroup prevalence scenario, the conventional FH (0, 1) generally yielded the highest power when hazard functions cross, in other words when the hazard ratio of the main treatment group is larger than 1. Similarly to the nTS tests, some of the pTS tests manage to improve on the power of the conventional tests when the survival probability of the treatment group is strictly superior. However, the power gain is relatively small. In contrast, some of the pTS tests were favourable in scenarios of small effect sizes when the subgroup prevalence is high. We observed improvements over both the associated conventional methods only at the edge of the investigated domain and only for selected second-step tests under NPH. Predominantly, the power of the pTS tests again is a compromise between the ones of the associated conventional tests. In general, the modestly weighted tests performed the best in the high subgroup prevalence scenario and the Modest (6) was also favourable when the subgroup prevalence is low and survival probability of the experimental group is strictly superior.

\FloatBarrier

\FloatBarrier

\section{Case study: Nivolumab in patients with non–small-cell lung cancer}

In addition to simulated scenarios, we applied the pTS method to analyze a reconstructed real-world data set from of a clinical trial that investigated the efficacy of Nivolumab in patients with nonsquamous non–small-cell lung cancer \cite{borghaei2015nivolumab}. Randomization to the experimental arm (with Nivolumab) and the control group (treated with Docetaxel) was done stratified according to prior maintenance treatment and line of therapy. Treatment was discontinued after disease progression events or adverse effects of the drug. The primary endpoint for the analysis was overall survival, progression-free survival was investigated as a secondary outcome.

Focusing on the secondary outcome of progression-free survival, we digitized the KM-estimates via the WebPlotDigitizer \cite{rohatgi2022} and applied Guyot´s method \cite{guyot2012enhanced} for the data reconstruction. The estimated curves are displayed in the introduction sections (see Figure \ref{nivolumab}).

In the actual trial, the authors tested for differences in survival probability with a log-rank test stratified analogous to the randomization criteria. The hazard ratio and its confidence interval was estimated with a stratified Cox model. The Nivolumab group was inferior in terms of both overall and progression-free survival in the first months of the study period. At some point the KM-Meier curves crossed and the experimental treatment showed long-term benefits. In case of PFS, the crossing happened around 7 months within the observation period. In the original paper, the pre-planned stratified log-rank test for PFS did not yield statistical significance.

We considered the same pre-test significance levels of $\alpha_{pre} \in \{0.05, 0.2\}$ and $m = 10000$ treatment label permutations in the two-step framework. Aside from that, we used the same conventional procedures as in the simulation study before. As before, all tests were one-sided at significance level $\alpha = 0.025$. The results are displayed in Table \ref{tab:res_case}.

\begin{table}[!ht]
    \centering
    \begin{tabular}{|p{1.6in}|>{\centering\arraybackslash}p{1.5in}|>{\centering\arraybackslash}p{1.5in}|>{\centering\arraybackslash}p{1.5in}|}
   \hline
   \textbf{Method} & \multicolumn{3}{c|}{\textbf{p-value to check PH assumption}} \\
    \hline
    GT test &   \multicolumn{3}{c|}{   $< 0.0001$}         \\   
    \hline
      & \multicolumn{3}{c|}{\textbf{p-value to check for treatment effect}} \\
      \hline
       \textbf{Method} & Conventional test & pTS test ($\alpha_{pre} = 0.05)$ & pTS test ($\alpha_{pre} = 0.2)$ \\
      \hline
     
      Log-rank test &   $0.1336$   & -&    -     \\
      FH (0, 1) &     $< 0.0001$ &  $0.0001$ & $< 0.0001$ \\
      FH (1, 1) &       $0.1091$ &  $0.1154$ &   $0.1104$ \\
      FH (1, 0) &       $0.9501$ &  $0.9467$ &   $0.9400$ \\
      Max-combo &     $< 0.0001$ &  $0.0001$ & $< 0.0001$ \\
      Modest (6) &      $0.0003$ &  $0.0004$ &   $0.0002$ \\
      Modest (12) &   $< 0.0001$ &  $0.0001$ & $< 0.0001$ \\
      \hline
    \end{tabular}
    \caption{\textbf{Case study:} P-values of the (weighted) log-rank tests and Max-combo test as well as their respective permutation two-step ("pTS") approaches with significance levels $\alpha_{pre} \in \{0.05, 0.2\}$ when analyzing the reconstructed data on progression-free survival in the Nivolumab-trial of advanced non-small lung cancer. The column "Conventional test" displays the p-value of the unconditional method denoted in the first column. For the two-step test either a log-rank test or the respective test of the first column is taken based on the results of the GT test to check the PH assumption. The adjusted p-values of the pTS tests are presented (the p-value of the naive two-step ("nTS") test would correspond to the the p-value shown in column 2). }
    \label{tab:res_case}
\end{table}

The unweighted log-rank test based on reconstructed data yielded a p-value of 0.1336 and would therefore not reject the null-hypothesis in favour of the experimental treatment. Weighted tests with late emphasis and the Max-combo test on the other hand recognize the long-term benefit of Nivolumab with p-values of $< 0.0001$, except the Modest (6), which gave a p-value of 0.0003.

P-values of the pTS tests show only subtle differences compared to the ones of the associated alternative tests under NPH. This was to be expected as the p-value of the GT-test is very small in this example. Consequently, the alternative test is performed in case of both pre-test significance levels. Two-step tests with alternatives that consider late effects also consistently suggest overall benefit of the experimental treatment with p-values of $\approx 10^{-4}$ or lower. Weighted tests with earlier emphasis, namely the FH (1, 0) and the FH (1, 1) fail to reject the null-hypothesis. The same was true for the pTS tests with the aforementioned tests as alternatives under NPH.

In the original trial, the estimated hazard ratio for PFS of the stratified Cox model was 0.92 ($p = 0.39$), which indicates a non-significant treatment effect. In this reconstructed example, the unweighted log-rank test similarly did not reject the null-hypothesis of identical survival distributions. However, tests with late emphasis reveal the long-term positive impact. Accounting for the possibility of non-proportional hazards via the two-step method or a Max-combo test also would have resulted in statistically significant p-values for comparing PFS.

It is important to note that we merely aimed to reconstruct the PFS curves of the respective study groups without consideration of associated covariates in order to obtain a representation of a real-life scenario of non-proportional hazards. Therefore, we do not claim that we could meaningfully analyze the trials outcome in any way and our results should not be mistaken for or misunderstood as actual outcomes of the study.

\section{Discussion}

The issues of using the log-rank test and the Cox proportional hazards model in the presence of non-proportional hazards have long been recognized. While many sources agree that departing from these two methods under these circumstances is generally advisable, as for now there is no consensus about best practices although numerous methods to analyze TTE-data have been suggested. In the methods sections we briefly discussed the idea of modifying the log-rank test. It has been established that with a reasonable understanding of treatment effects prior to the trial, appropriately weighting observations according to their time in the follow-up period can improve on the log-rank test´s power under non-proportional hazards. On the other hand, faulty assumptions can lead to the opposite effect. In instances, where many hazard functions may be considered reasonable, combination tests provide a robust alternative.

The considered two-stage method can be seen as a compromise between fully sticking to the log-rank test or always using an alternative test instead. Based on the observed time-to-event data the pre-test will determine whether it is reasonable to stay with the familiar log-rank test or not. The considered method might be a reasonable backup strategy when there are some uncertainties whether the PH assumptions is reasonable. For example, if there are strong assumptions that the PH applies, then a classical sample size calculation for the log-rank test could be performed, but using the pTS as the primary test. Clinical trial simulations can be used to evaluate how the power is affected for other NPH scenarios. If there is stronger evidence for NPH, then the sample size has to be determined by clinical trial simulations for the NPH scenario(s) of interest.

We modified the two-step procedure to be able to control the type-I error when the hazard functions are identical. We argue that this works similar to other permutation tests in settings of exchangeability and illustrated a working example in the supplementary material. As the permutation distribution is generated conditionally on the survival function of the pooled observations, it corresponds to a null hypothesis of equal survival curves. Therefore, the permutation test might not control the type-I error rate in scenarios where the survival curve under treatment is strictly dominated by the survival curve under control, but the hazard functions cross \cite{magirr2019modestly}.

Here, we would also like to recognize other robust tests for comparing survival probabilities under NPH that share some similarities with our work. Royston and Parmar suggested a joint test that combines the log-rank and GT test statistics, which, according to their simulation study, outperforms the standard log-rank test in scenarios involving delayed or diminishing effects, at some power cost under proportional hazards \cite{royston2014approach}. In another work, they introduced a combined test of the standard Cox PH test and an RMST-based permutation test \cite{royston2016augmenting}. They found that similar to the joint test it sacrifices some power under PH, but remains advantageous in cases of delayed and diminishing treatment effects, making it a "recommendable as an omnibus test of a generalized treatment effect". Qiu and Sheng proposed a two-step testing approach that first employs a log-rank test and then follows up with a test designed to address issues arising when hazards cross, should the log-rank test not reject \cite{qiu2008two}. This method resolves the challenges faced by log-rank tests in scenarios of crossing survival curves.

A similar framework to our work was introduced by Campbell and Dean \cite{campbell2014consequences}, whereby a Cox regression based test is suggested under PH. Under NPH they examined as second-step tests time-dependent Cox PH models with different trends, a conditional log-rank test \cite{logan2008comparing} and accelerating failure time (AFT) models \cite{wei1992accelerated}. In addition to the top-down approach they investigated a permutation adjustment conditioned on the original result of the GT-test. They concluded that the approach might be used "when the nature of the treatment effect cannot be determined with a high degree of confidence before obtaining the data". In a more extensive simulation study, Callegaro et al. \cite{callegaro2017testing} found that the adjusted two-step test with a time-dependent Cox-model as an alternative test under NPH, while robust, lacks power to "optimal" tests under different scenarios. 

We investigated the two-step approach with variations of the log-rank test for the comparison of the time-to-event outcomes and observed similar results. Under non-proportional hazards, the permutation tests proved generally favourable to the log-rank test if the alternative test under NPH is chosen appropriately. However, this came at the expense of reduced power when the PH assumption was met. In selected scenarios, we observed that the permutation two-step tests could surpass both the associated conventional methods. Although we did not cover this aspect in the results section, it is important to mention that the performance of the whole two-step procedure is closely linked to the pre-test significance level. This is an aspect that has not been addressed in previous works, as they considered pre-test significance levels of $\alpha_{pre} = 0.05$ or lower. When higher values are considered, it increases the likelihood of using the alternative method under non-proportional hazards. As the trial is not powered for the pre-test, using higher significance levels $\alpha_{pre}$ is preferable the higher the uncertainty with regards to the applicability of the PH assumption is. The ability to adjust the method might be an advantage as it enables researchers to incorporate their expectations into the decision-making process.

In this paper we focused on the operating characteristics of the underlying testing procedure. An open research problem is which treatment effect should be actually reported. If the PH assumption holds, comparing the hazard functions by means of the hazard ratio will be the natural choice. For NPH, it is not obvious which alternative should be considered to compare survival probabilities between two groups. One alternative suggested is the restricted mean survival time (RMST), an estimator for the area under the survival curve up to some time point $t^*$. Its advantage is that it is easily interpretable for the clinical community. On the downside, tests based on RMST suffer from little power in scenarios of late treatment effects or crossing hazard functions \cite{lin2020alternative} \cite{klinglmuller2023neutral}. Furthermore, the performance depends on the choice of the hyperparameter $t^*$. Point estimators such as the median survival or survival probabilities at specific time points may also be used. Both of these have the disadvantage of reducing the information from survival curves to a single point. In the case of milestone survivals, a mis-specification of the time point can also lead to significant power loss \cite{ristl2021delayed}. In a NPH setting it seems reasonable to not select only one, but report more than one of the parameters above, \citep{ristl2023simultaneous} suggested to use simultaneous tests allowing for simultaneous confidence intervals for different metrics of interest, e.g. for differences or ratios of milestone survival probabilities or quantiles, differences in RMST and an average hazard ratio. For a more comprehensive overview of methods for analyzing time-to-event data under non-proportional hazards, we refer to \cite{bardo2023methods}. Subject to future research is also how to derive group sequential testing boundaries for the two-stage test similar to \cite{li2023group, ghosh2022robust}. 

Finally, we would like to stress that having a precise understanding of the mechanism of action and the resulting survival curves is generally preferable than applying a non-specific method that, while robust to many deviations from PH, will be less informative with respect to the suspected causes for NPH and  potentially also less powerful than methods tailored to the specific scenario. For well-known patterns of non-proportional hazards suitable (often even highly scenario-specific) methods have been proposed. Similar to the max-combo test, the two-stage test could be a useful alternative in instances where the mechanism of action cannot be precisely determined a priori.

\clearpage

\bibliographystyle{unsrtnat}
\bibliography{bibliography}

\clearpage

\appendix

\section{Permutation two-step test as an algorithm} \label{alg:random}

The following algorithm is to perform the permutation two-step (pTS) test for two-group comparison of time-to event data in randomized clinical trials from section 3 in the main material. List of abbreviations:

\begin{itemize}
\item TTE \ldots Time-to-event
    \item PH \ldots Proportional hazards
    \item GT-test \ldots Grambsch-Therneau test
    \item LR-test \ldots Log-rank test
\end{itemize}

\begin{algorithm}
\textbf{Require:} Alternative test after rejection of the PH-assumption\\
\textbf{Require:} $\alpha \in [0, 1] \ldots$ significance level for comparison of TTE-endpoints \\
\textbf{Require:} $\alpha_{pre} \in [0, 1] \ldots$ significance level for pre-test of PH \\
\textbf{Require:} $m > 0 \ldots$ Number of permutations of the treatment label \\
$k \gets 0$\\
$p \gets 0$\\
$p_{pre} \gets $ p-value of GT-test \\
\eIf{$p_{pre} > \alpha_{pre}$}{
$p_0 \gets $ p-value of the LR-test
}{$p_0 \gets $ p-value of the alternative test}
\While{$k < m$}{
Randomly permute treatment covariate\\
\eIf{$p_{pre} > \alpha_{pre}$}{
$p_1 \gets $ p-value of the LR-test of permuted data
}{$p_1 \gets $ p-value of the alternative test of permuted data}
$p \gets p + \frac{\mathbbm{1}_{p_1 \leq p_0}}{m}$\\
$k \gets k + 1$
}
\Return $p \leq \alpha$
\end{algorithm}

\FloatBarrier
\clearpage

\setcounter{figure}{0}
\makeatletter 
\renewcommand{\thefigure}{A\@arabic\c@figure}
\makeatother

\section{Simulation scenarios}

\begin{figure}[!ht]
\centering
\includegraphics[width=.85\textwidth]{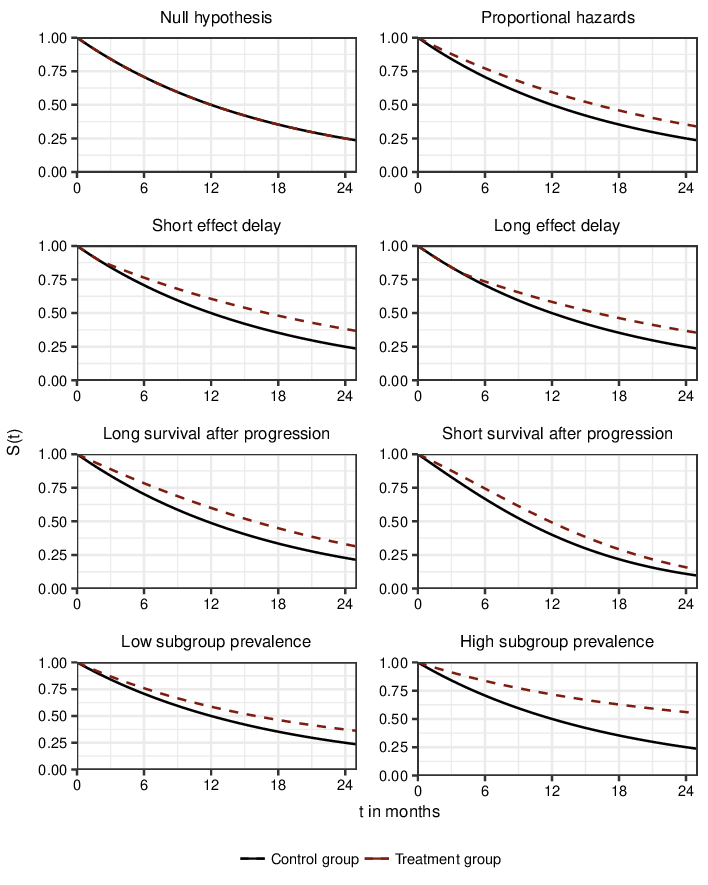}
\caption{\textbf{Simulation study:} Survival curves of the study groups under the reference scenarios. The hazard rate of the treatment group as a whole, after effect onset, without disease progression events or in the subgroup-complement (depending on the scenario) is varied within the simulations.}
\label{scenarios}
\end{figure}

\FloatBarrier
\clearpage

\section{Results of the effect delay and disease progression scenarios}

\begin{figure}[H]
	\makebox[\textwidth][c]{\includegraphics[width=.9\textwidth]{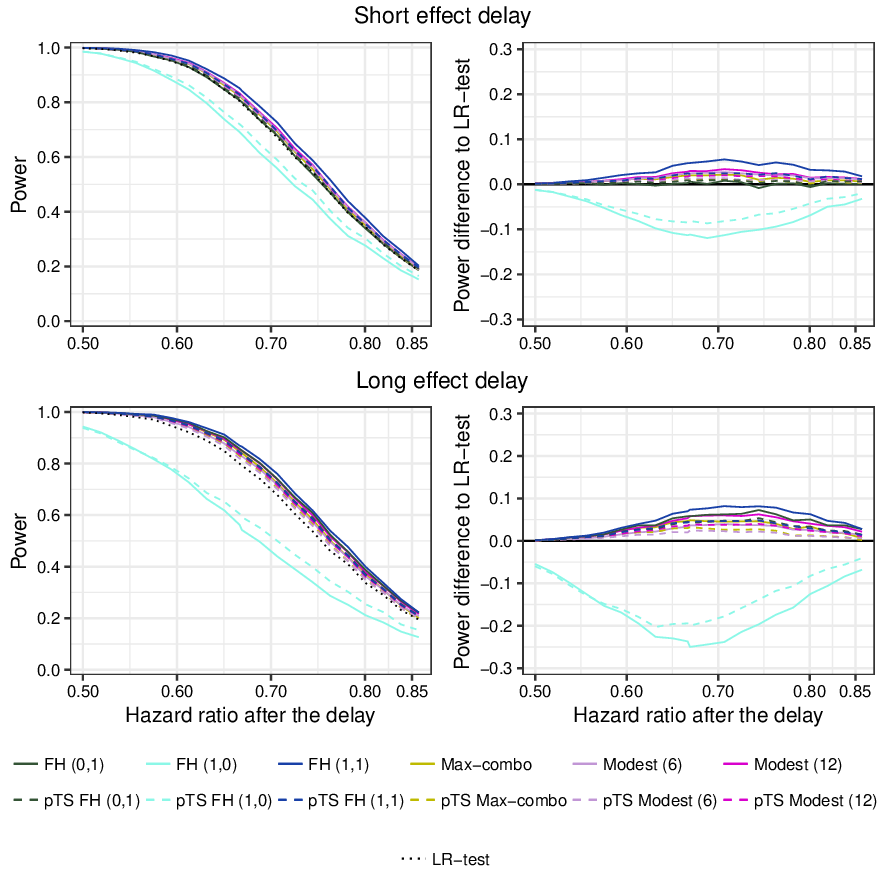}}
 \caption{\textbf{Delayed effect scenarios:} Absolute power (left column) and power difference (right column) to the log-rank test of the permutation two-step ("pTS") tests. Results of the two-step tests are displayed as dashed lines in the same colour as the associated conventional approaches under NPH. The power of the log-rank test ("LR-test") serves as a reference and is displayed as a black dotted line.}
	\label{res:randomdelay}
\end{figure}

\begin{figure}[H]
	\makebox[\textwidth][c]{\includegraphics[width=.9\textwidth]{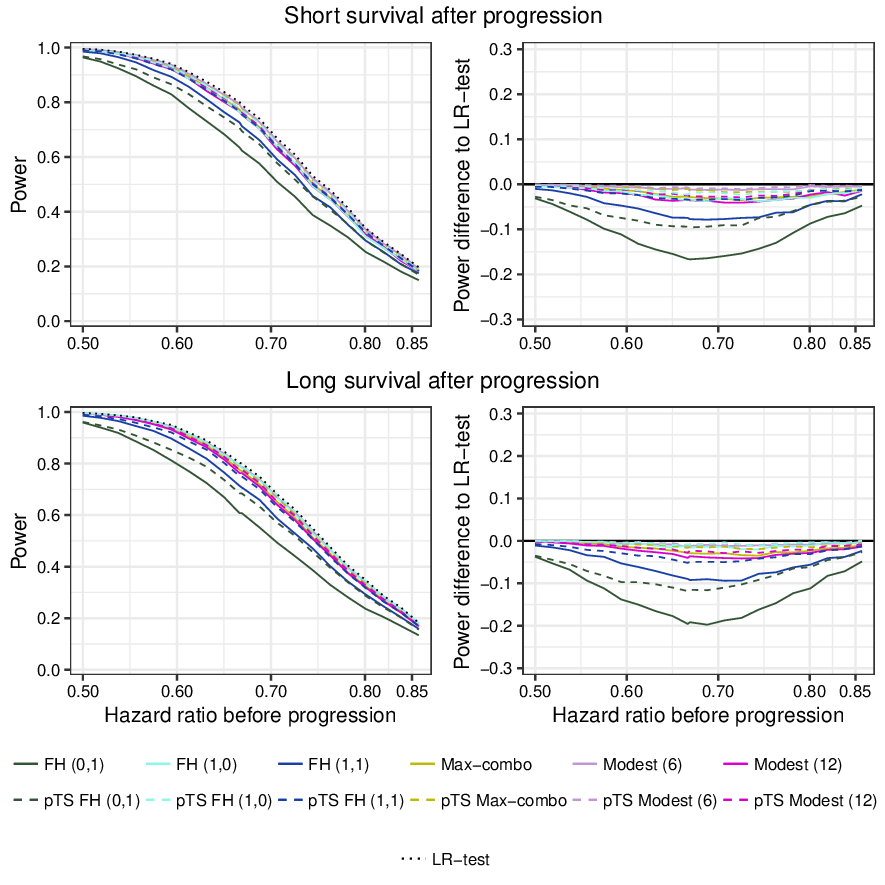}}
 \caption{\textbf{Disease progression scenarios:} Absolute power (left column) and power difference (right column) to the log-rank test of the permutation two-step ("pTS") tests. Results of the two-step tests are displayed as dashed lines in the same colour as the associated conventional approaches under NPH. The power of the log-rank test ("LR-test") serves as a reference and is displayed as a black dotted line.}
	\label{res:randomprogression}
\end{figure}

\end{document}


\title{Supplementary material for: A two-step testing approach for comparing time-to-event data under non-proportional hazards}
\author{}
\date{}

\maketitle

This supplementary material of \textit{A two-step testing approach for comparing time-to-event data under non-proportional hazards} contains a power comparison of the permutation two-step (pTS) tests with different pre-test levels ($\alpha_{pre} \in \{0.05, 0.2\})$) and results of the naive two-step (nTS) approach in alternative scenarios of different sample size and recruitment speed.
The simulation setup is explained in section 4.1 of the main manuscript containing a detailed description of the different scenarios. An overview of different design choices and assumptions investigated are given in Table 1 of the main paper.

\setcounter{figure}{0}
\makeatletter 
\renewcommand{\thefigure}{S\@arabic\c@figure}
\makeatother

\section{Performance measures of permutation two-step tests with regards to pre-test level}

\FloatBarrier

\begin{figure}[H]
\centering
\includegraphics[width=.87\textwidth]{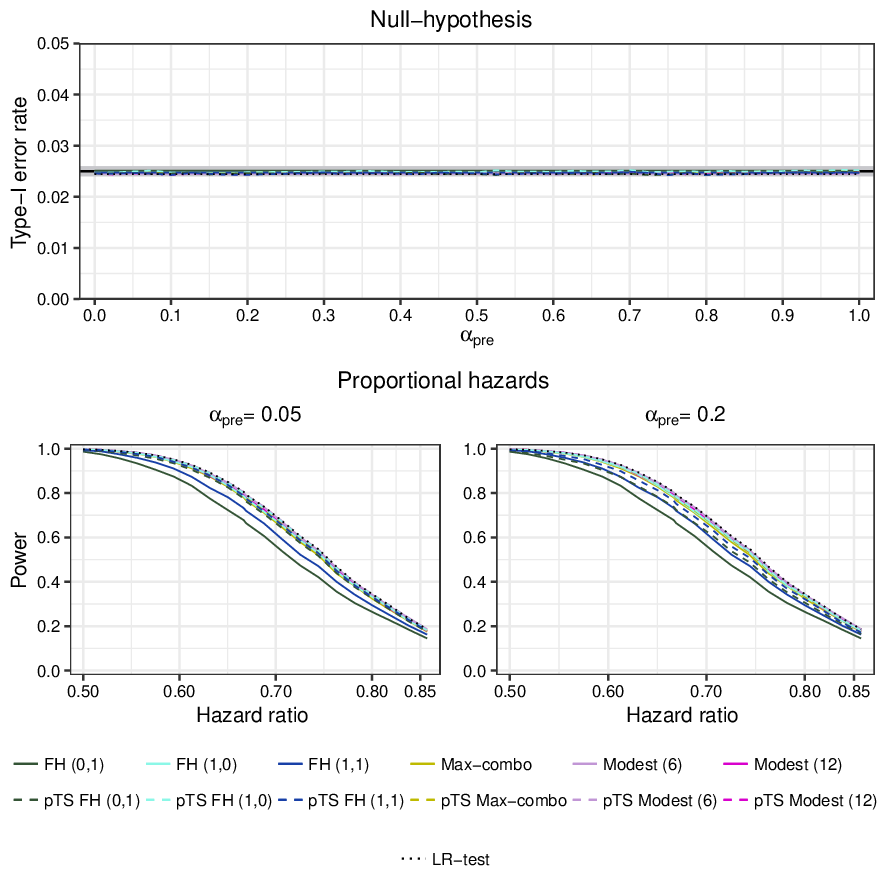}
\caption{\textbf{Null-hypothesis:} Type-I error rates of the permutation two-step ("pTS") tests. The 95\% prediction interval of the estimated type-I error rate under identical hazard functions is marked in grey. The one-sided significance level $\alpha = 0.025$ is indicated as a solid black line. \textbf{Proportional hazards scenario:} Power of the pTS tests with pre-test significance levels of $\alpha_{pre} = 0.05$ (left column) and $\alpha_{pre} = 0.2$ (right column). Results of the two-step tests are displayed as dashed lines in the same colour as the respective unconditional approaches under NPH. The power of the log-rank test ("LR-test") serves as a reference and is displayed as a black dotted line.}
\label{res:nullidentical_supp}
\end{figure}

\begin{figure}[H]
	\makebox[\textwidth][c]{\includegraphics[width=.87\textwidth]{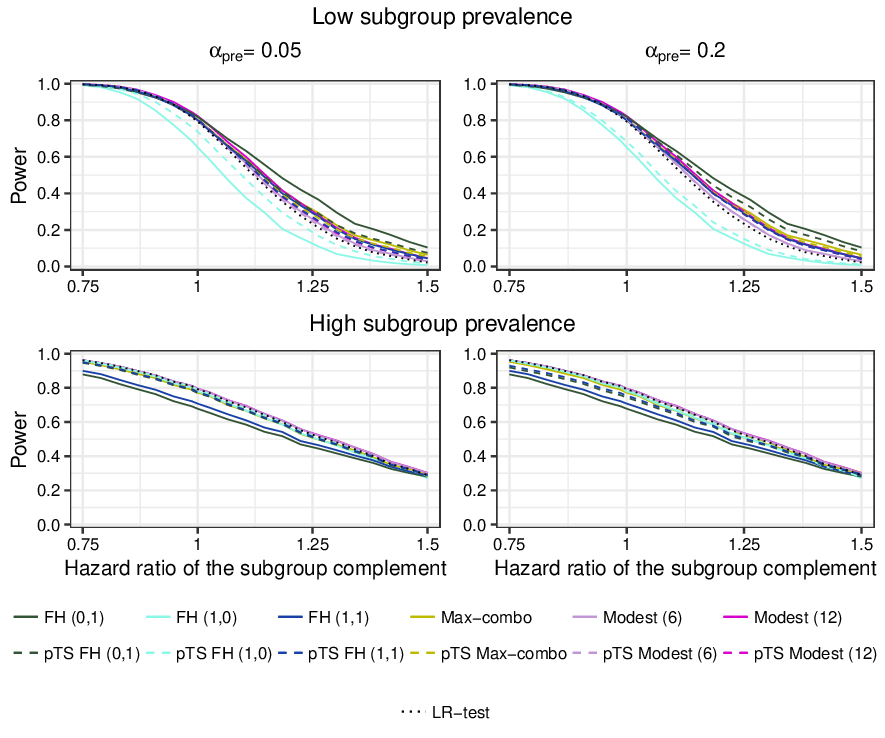}}
 \caption{\textbf{Subgroup scenarios:} Power of the permutation two-step ("pTS") tests with pre-test significance levels of $\alpha_{pre} = 0.05$ (left column) and $\alpha_{pre} = 0.2$ (right column). Results of the two-step tests are displayed as dashed lines in the same colour as the respective unconditional approaches under NPH. The power of the log-rank test ("LR-test") serves as a reference and is displayed as a black dotted line. The hazard ratio is with regard to the complement of the treatment subgroup. The HR of the treatment subgroup compared to the control is always 0.1 (corresponding to a median survival time of 120 months).}
	\label{res:randomsub_appen}
\end{figure}

\begin{figure}[H]
	\makebox[\textwidth][c]{\includegraphics[width=.87\textwidth]{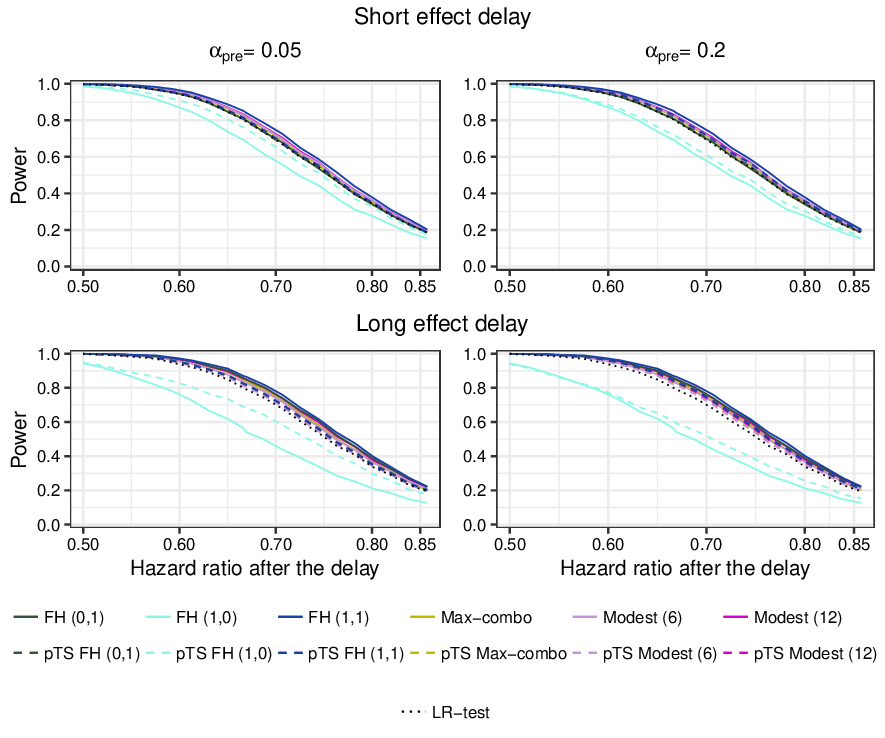}}
 \caption{\textbf{Delayed effect scenarios:} Power of the permutation two-step ("pTS") tests with pre-test significance levels of $\alpha_{pre} = 0.05$ (left column) and $\alpha_{pre} = 0.2$ (right column). Results of the two-step tests are displayed as dashed lines in the same colour as the respective unconditional approaches under NPH. The power of the log-rank test ("LR-test") serves as a reference and is displayed as a black dotted line.}
	\label{res:randomdelay_appen}
\end{figure}

\begin{figure}[H]
	\makebox[\textwidth][c]{\includegraphics[width=.87\textwidth]{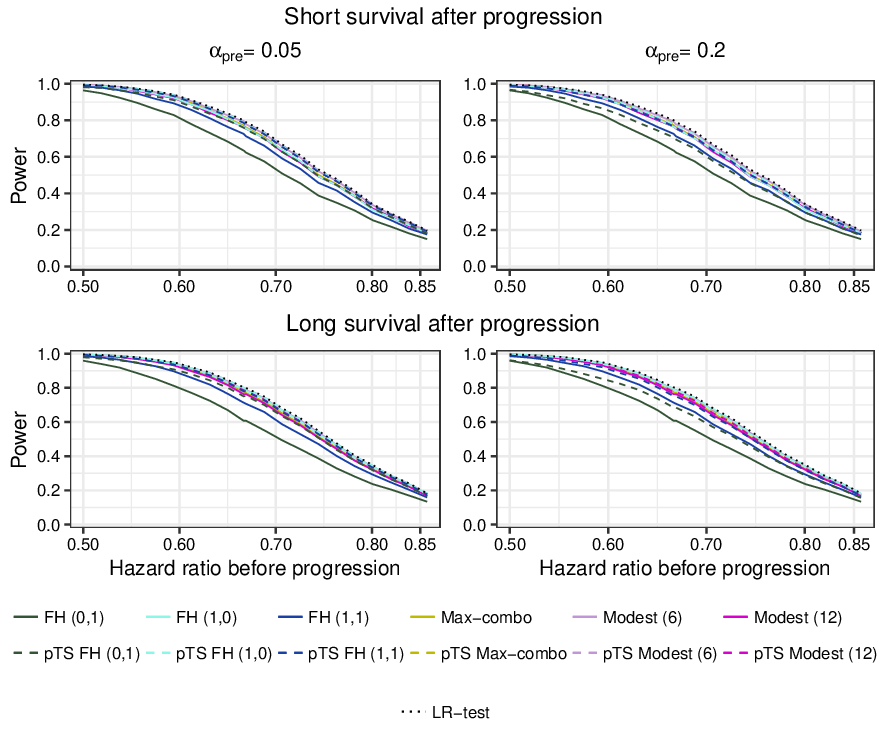}}
 \caption{\textbf{Disease progression scenarios:} Power of the permutation two-step ("pTS") tests with pre-test significance levels of $\alpha_{pre} = 0.05$ (left column) and $\alpha_{pre} = 0.2$ (right column). Results of the two-step tests are displayed as dashed lines in the same colour as the respective unconditional approaches under NPH. The power of the log-rank test ("LR-test") serves as a reference and is displayed as a black dotted line.}
	\label{res:randomprogression_appen}
\end{figure}

\clearpage

\section{Performance measures of the naive two-step test in additional scenarios}

\begin{figure}[H]
	\centering
  \includegraphics[width=.87\textwidth]{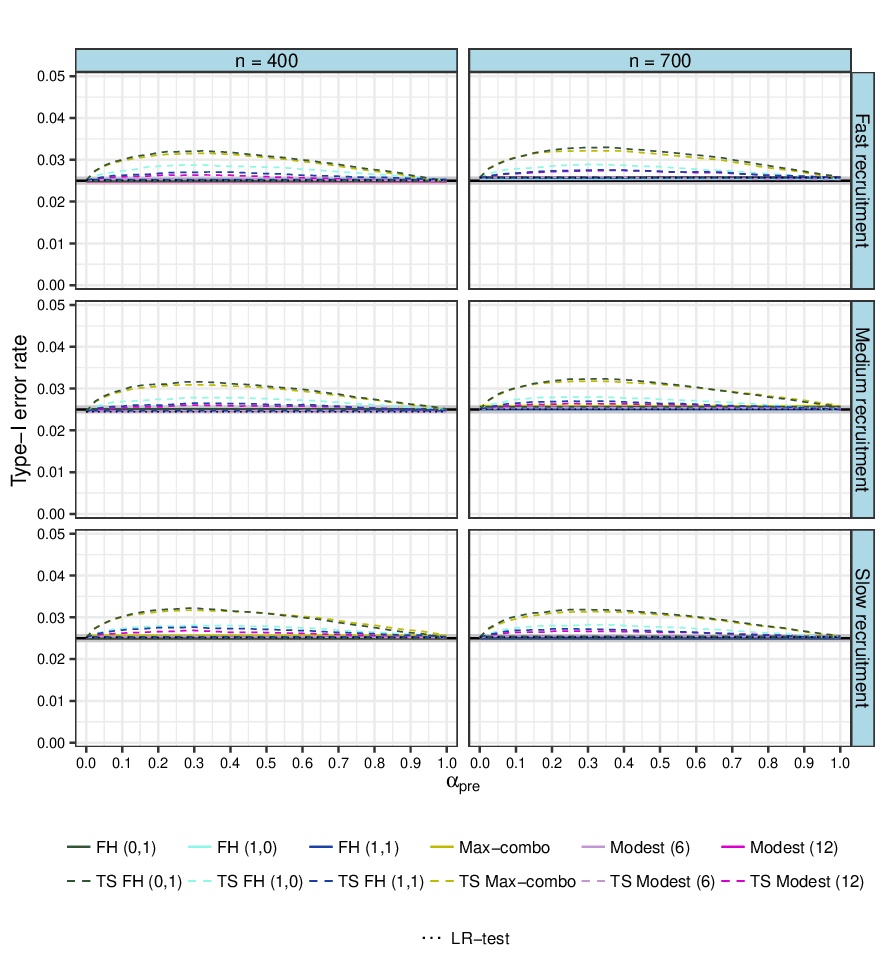}
 \caption{\textbf{Null-hypothesis:} Type-I error rates as function of the pre-test level. The 95\% prediction interval of the estimated type-I error rate under identical hazard functions is marked in grey. Values outside this area provide strong evidence of type-I error inflation. The one-sided significance level $\alpha = 0.025$ is indicated as a solid black line. The naive two-step test procedures ("nTS") are represented by dashed lines in the same colour as the associated conventional approaches. The results of the log-rank test ("LR-Test") are shown as a black dotted line.}
\end{figure}

\subsection{Power curves with $\alpha_{pre} = 0.05$}

\begin{figure}[H]
	\centering
  \includegraphics[width=.87\textwidth]{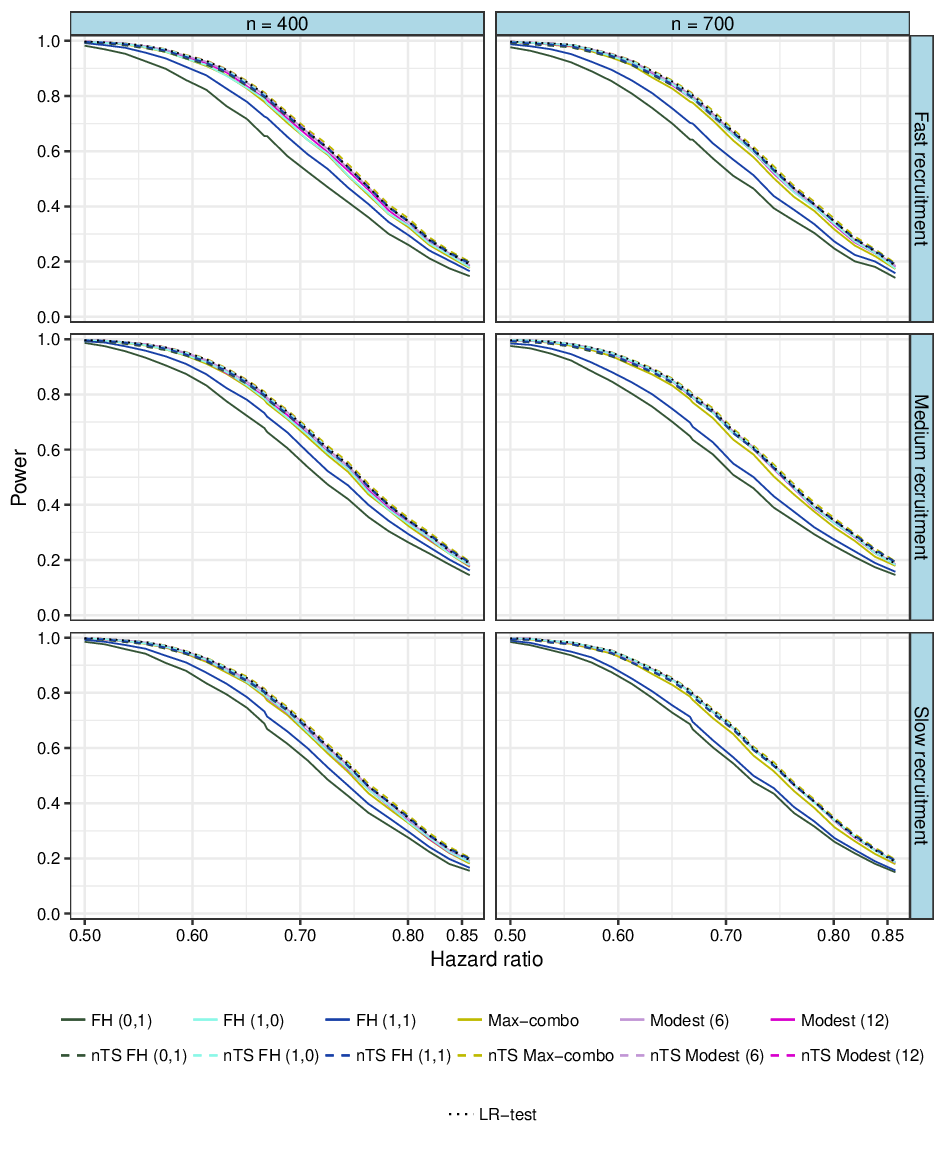}
 \caption{\textbf{Proportional hazards scenario:} Power of the naive two-step ("nTS") tests with a pre-test significance level of  $\alpha_{pre} = 0.05$. Results of the two-step tests are displayed as dashed lines in the same colour as the respective unconditional approaches under NPH. The power of the log-rank test ("LR-test") serves as a reference and is displayed as a black dotted line.}
\end{figure}

\begin{figure}[H]
	\centering
  \includegraphics[width=.87\textwidth]{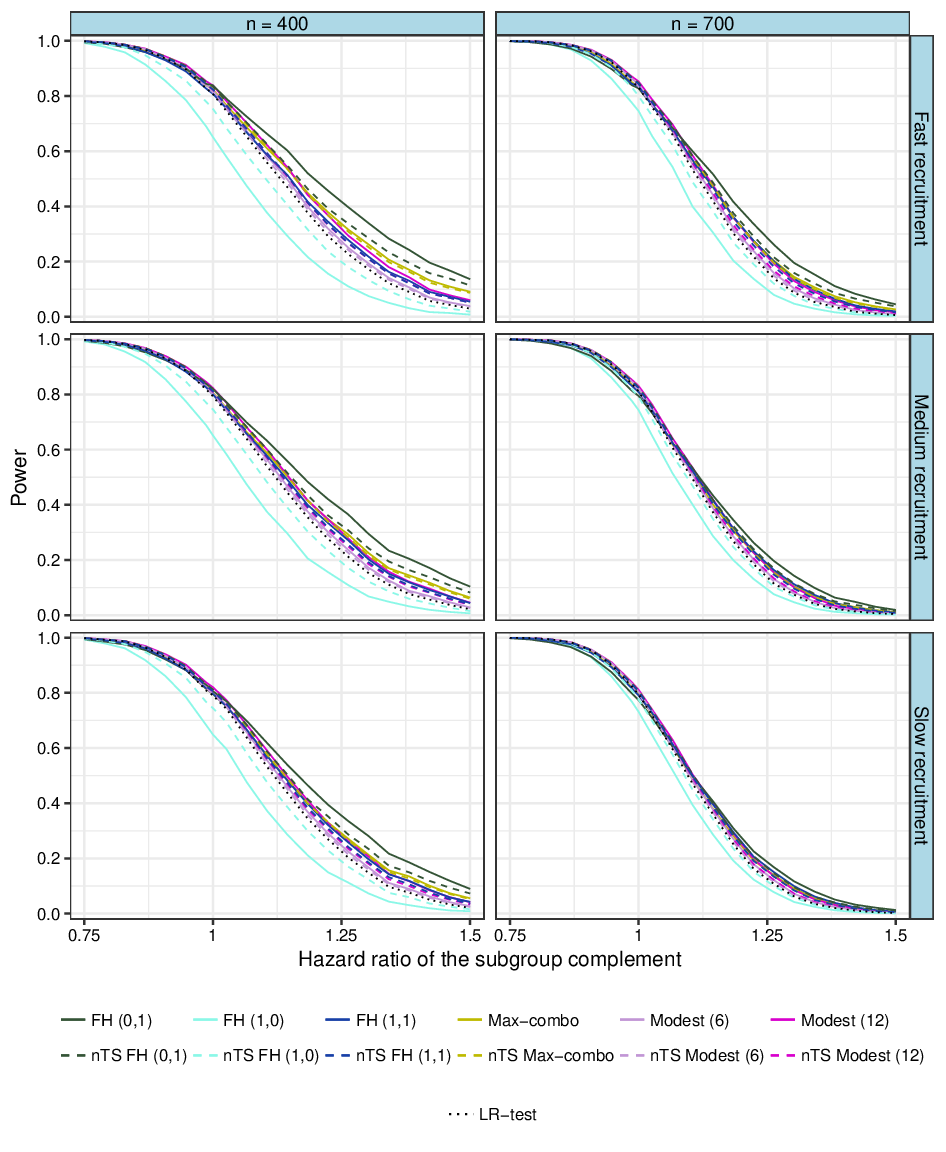}
 \caption{\textbf{Low subgroup prevalence scenario:} Power of the naive two-step ("nTS") tests with a pre-test significance level of  $\alpha_{pre} = 0.05$. Results of the two-step tests are displayed as dashed lines in the same colour as the respective unconditional approaches under NPH. The power of the log-rank test ("LR-test") serves as a reference and is displayed as a black dotted line. The hazard ratio is with regard to the complement of the treatment subgroup. The HR of the treatment subgroup compared to the control is always 0.1 (corresponding to a median survival time of 120 months).}
\end{figure}

\begin{figure}[H]
	\centering
  \includegraphics[width=.87\textwidth]{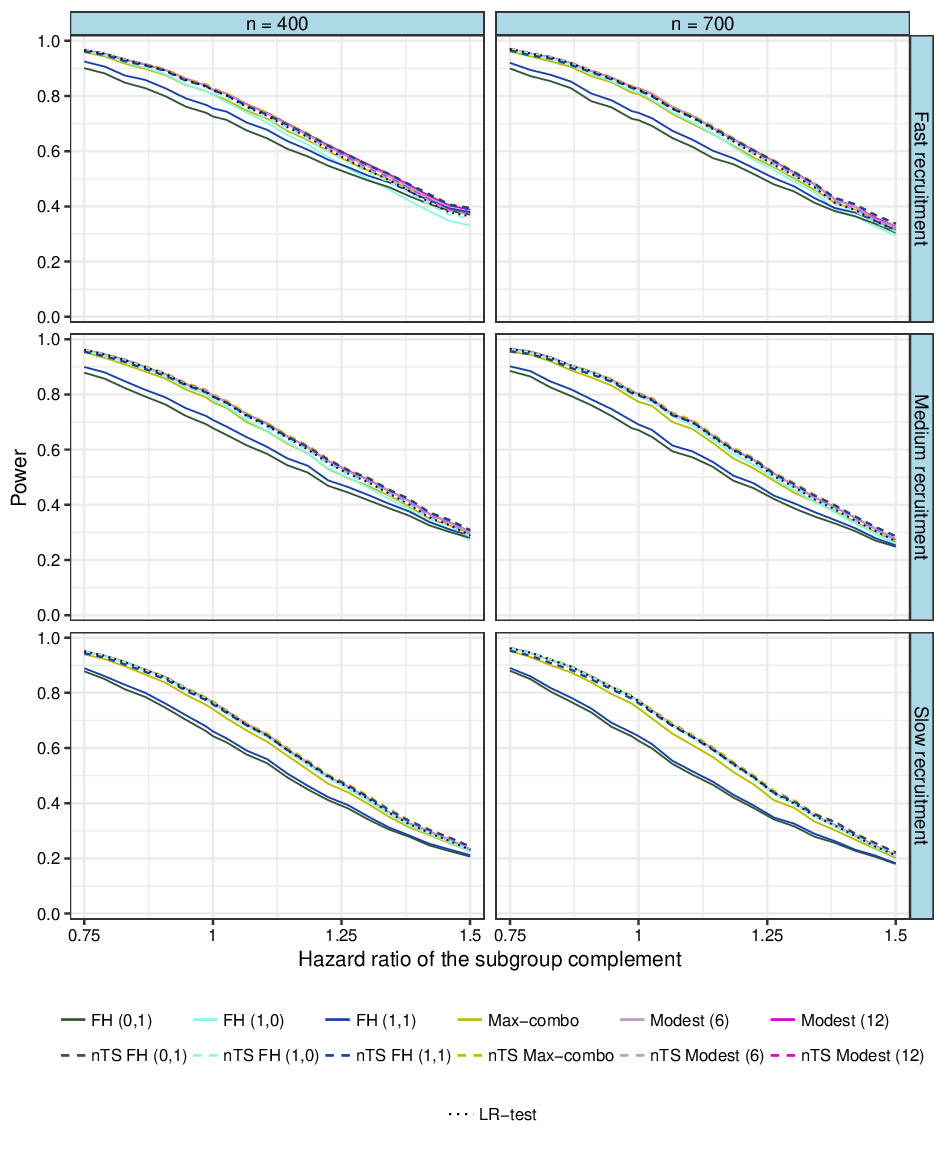}
 \caption{\textbf{High subgroup prevalence scenario:} Power of the naive two-step ("nTS") tests with a pre-test significance level of  $\alpha_{pre} = 0.05$. Results of the two-step tests are displayed as dashed lines in the same colour as the respective unconditional approaches under NPH. The power of the log-rank test ("LR-test") serves as a reference and is displayed as a black dotted line. The hazard ratio is with regard to the complement of the treatment subgroup. The HR of the treatment subgroup compared to the control is always 0.1 (corresponding to a median survival time of 120 months).}
\end{figure}

\begin{figure}[H]
	\centering
  \includegraphics[width=.87\textwidth]{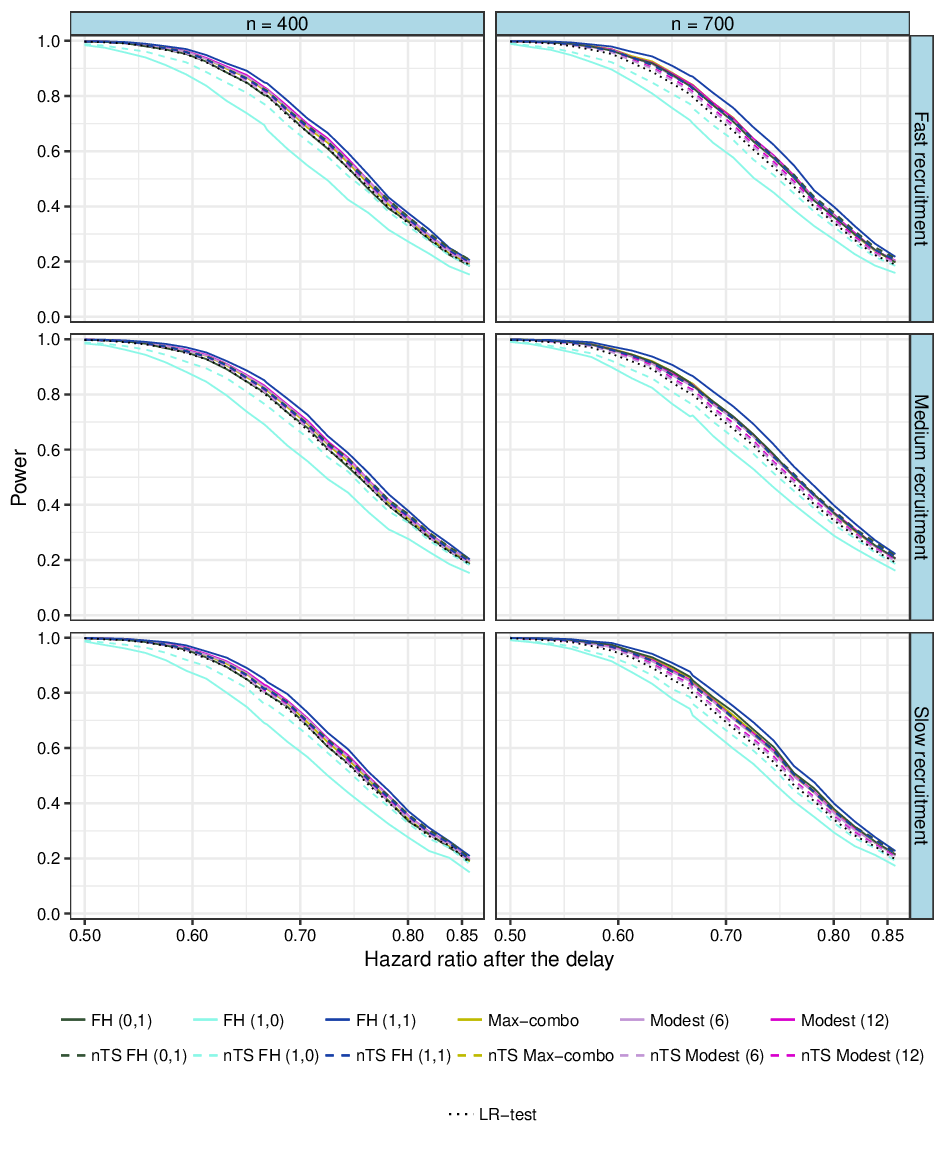}
 \caption{\textbf{Short delayed effect scenario:} Power of the naive two-step ("nTS") tests with a pre-test significance level of  $\alpha_{pre} = 0.05$. Results of the two-step tests are displayed as dashed lines in the same colour as the respective unconditional approaches under NPH. The power of the log-rank test ("LR-test") serves as a reference and is displayed as a black dotted line.}
\end{figure}

\begin{figure}[H]
	\centering
  \includegraphics[width=.87\textwidth]{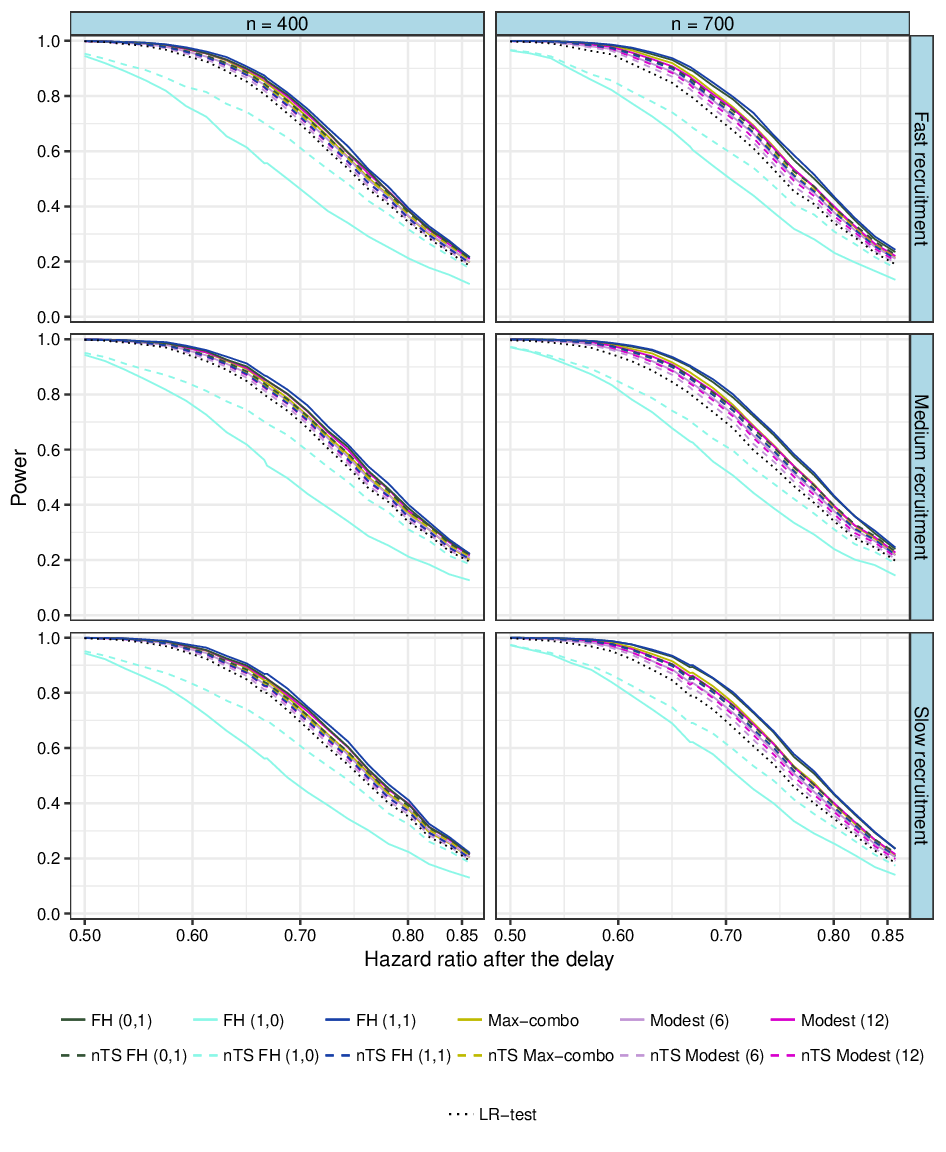}
 \caption{\textbf{Long delayed effect scenario:} Power of the naive two-step ("nTS") tests with a pre-test significance level of  $\alpha_{pre} = 0.05$. Results of the two-step tests are displayed as dashed lines in the same colour as the respective unconditional approaches under NPH. The power of the log-rank test ("LR-test") serves as a reference and is displayed as a black dotted line.}
\end{figure}

\begin{figure}[H]
	\centering
  \includegraphics[width=.87\textwidth]{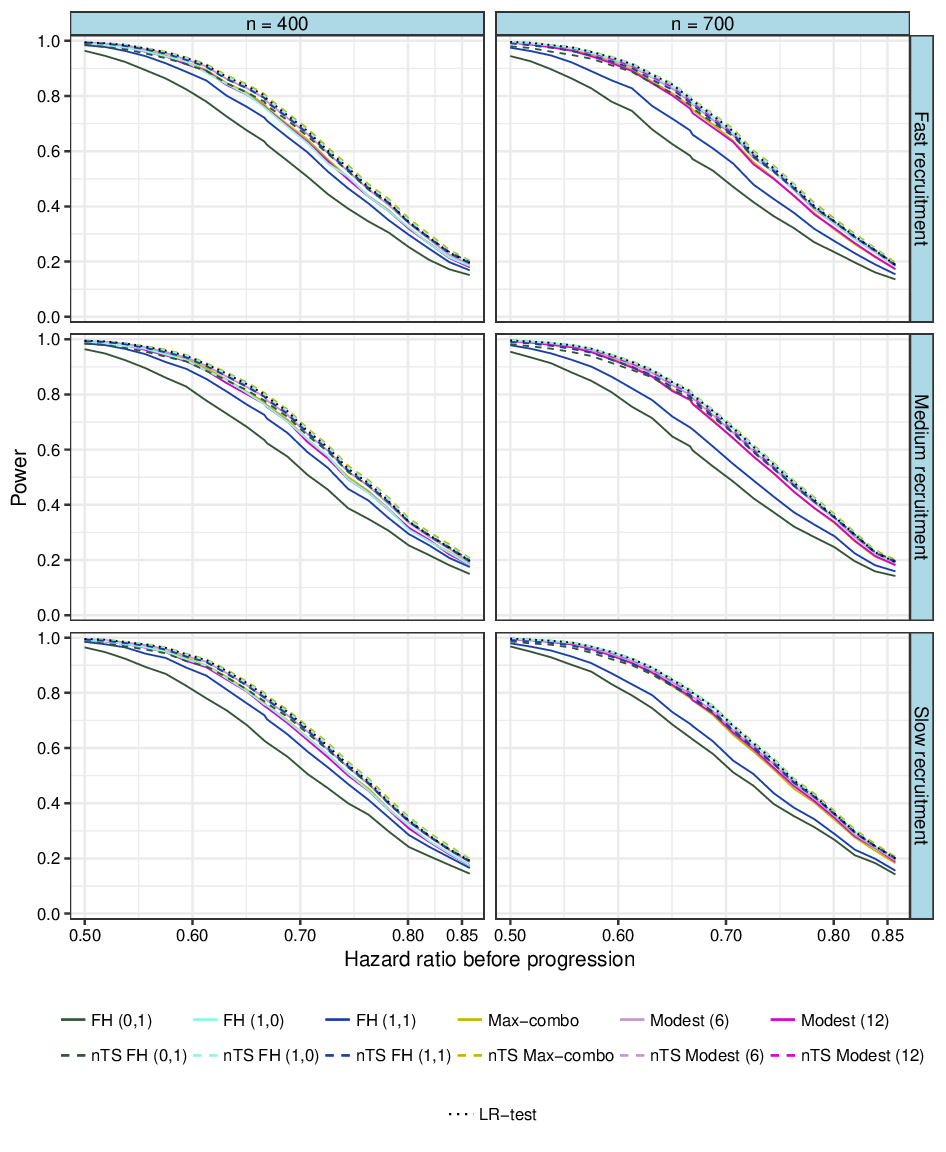}
 \caption{\textbf{Disease progression scenario (Short survival):} Power of the naive two-step ("nTS") tests with a pre-test significance level of  $\alpha_{pre} = 0.05$. Results of the two-step tests are displayed as dashed lines in the same colour as the respective unconditional approaches under NPH. The power of the log-rank test ("LR-test") serves as a reference and is displayed as a black dotted line.}
\end{figure}

\begin{figure}[H]
	\centering
  \includegraphics[width=.87\textwidth]{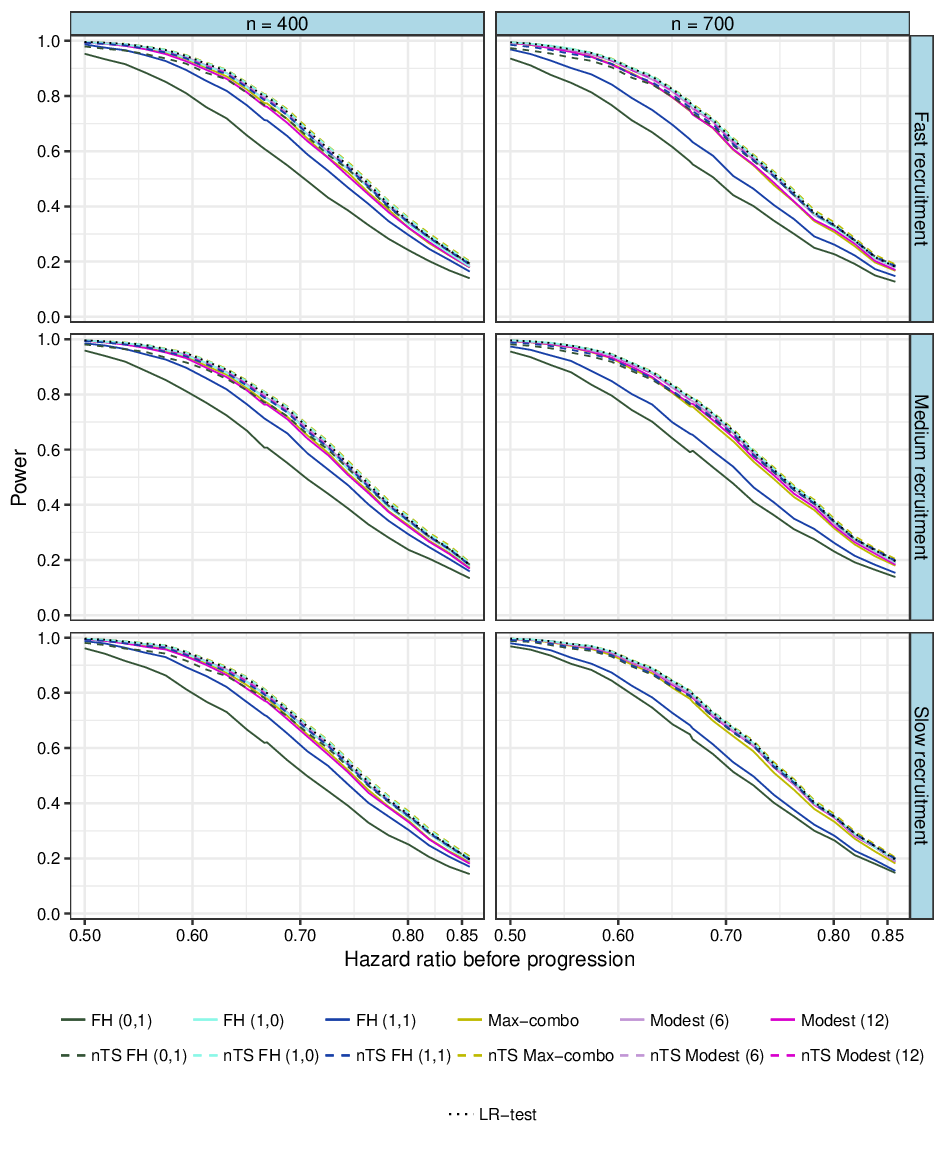}
 \caption{\textbf{Disease progression scenario (Long survival):} Power of the naive two-step ("nTS") tests with a pre-test significance level of  $\alpha_{pre} = 0.05$. Results of the two-step tests are displayed as dashed lines in the same colour as the respective unconditional approaches under NPH. The power of the log-rank test ("LR-test") serves as a reference and is displayed as a black dotted line.}
\end{figure}

\subsection{Power curves with $\alpha_{pre} = 0.2$}

\begin{figure}[H]
	\centering
  \includegraphics[width=.87\textwidth]{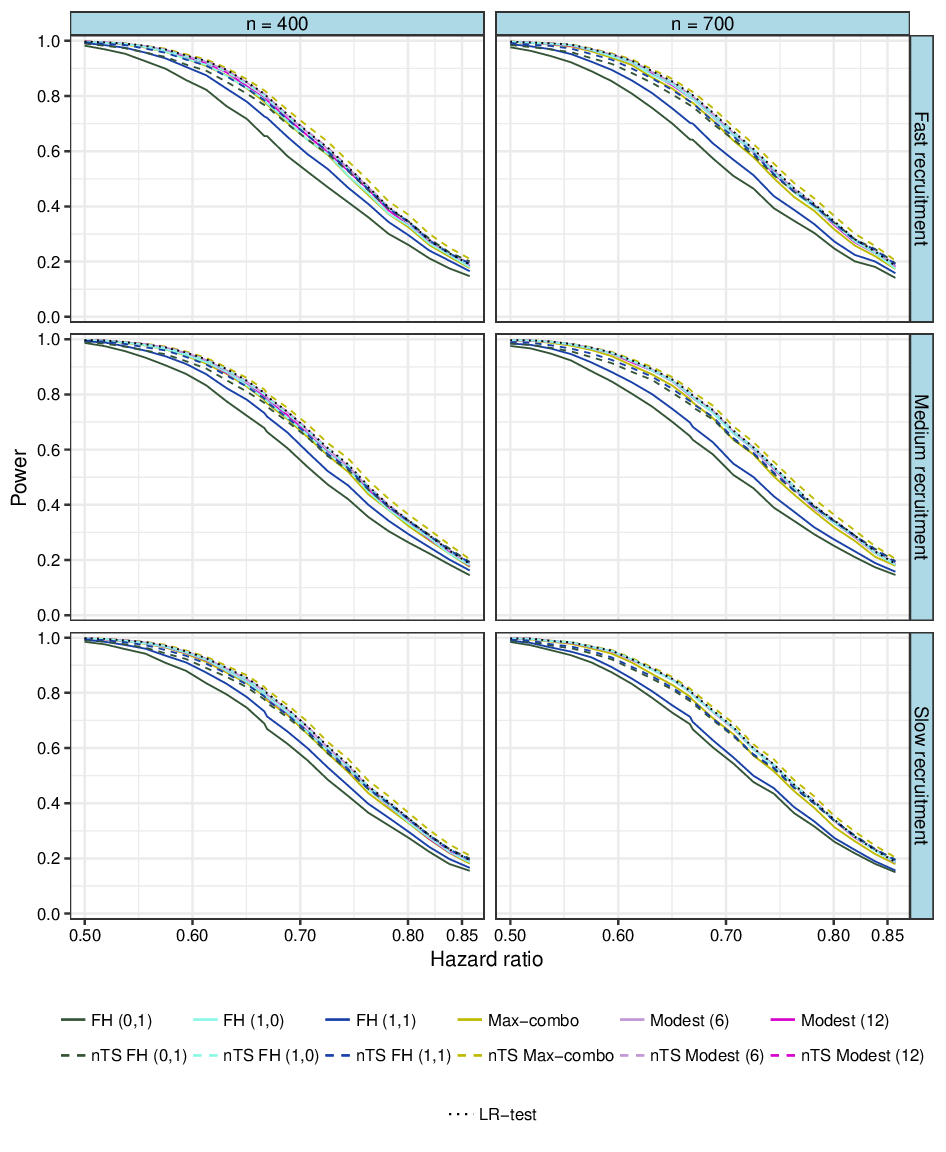}
 \caption{\textbf{Proportional hazards scenario:} Power of the naive two-step ("nTS") tests with a pre-test significance level of  $\alpha_{pre} = 0.2$. Results of the two-step tests are displayed as dashed lines in the same colour as the respective unconditional approaches under NPH. The power of the log-rank test ("LR-test") serves as a reference and is displayed as a black dotted line.}
	\label{res:ph_alpha0.2}
\end{figure}

\begin{figure}[H]
	\centering
  \includegraphics[width=.87\textwidth]{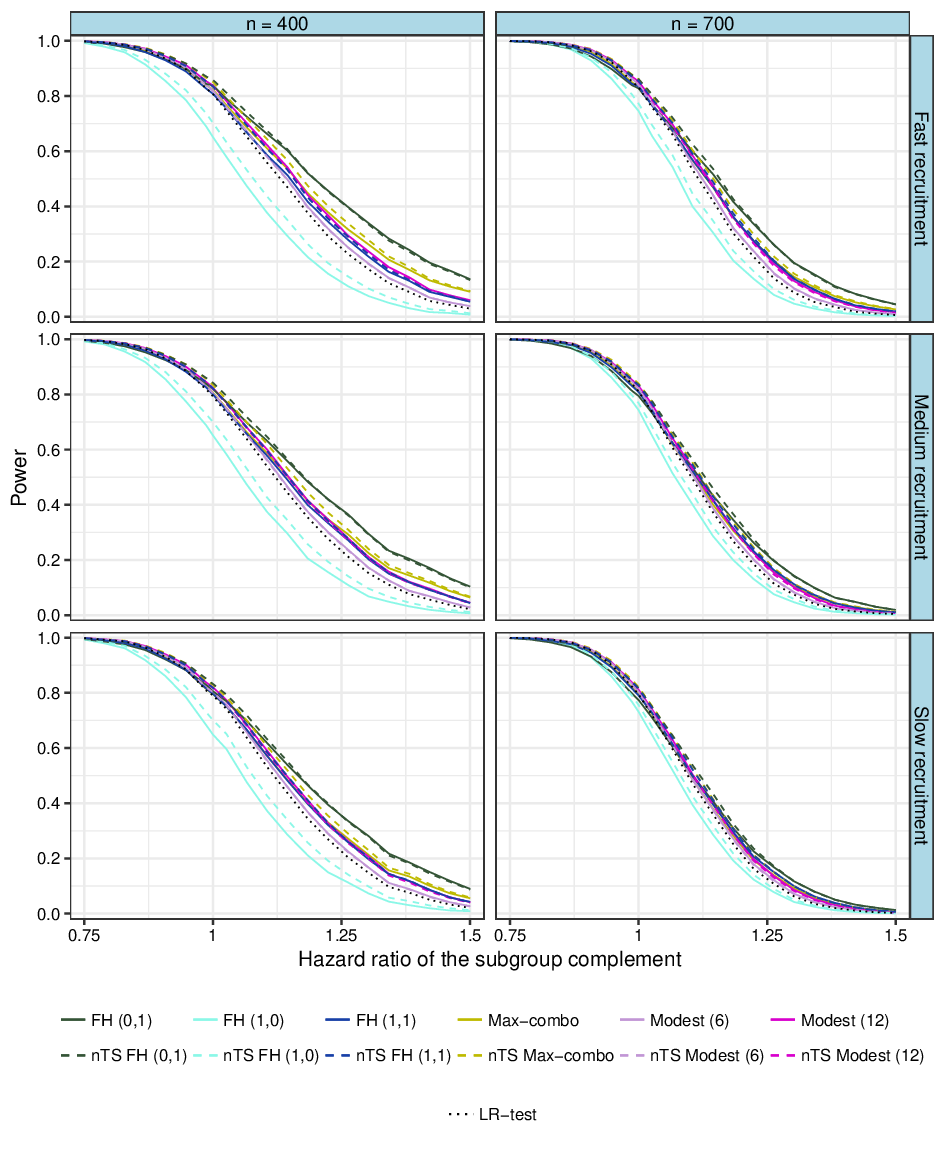}
 \caption{\textbf{Low subgroup prevalence scenario:} Power of the naive two-step ("nTS") tests with a pre-test significance level of  $\alpha_{pre} = 0.2$. Results of the two-step tests are displayed as dashed lines in the same colour as the respective unconditional approaches under NPH. The power of the log-rank test ("LR-test") serves as a reference and is displayed as a black dotted line. The hazard ratio is with regard to the complement of the treatment subgroup. The HR of the treatment subgroup compared to the control is always 0.1 (corresponding to a median survival time of 120 months).}
	\label{res:sub20_alpha0.2}
\end{figure}

\begin{figure}[H]
	\centering
  \includegraphics[width=.87\textwidth]{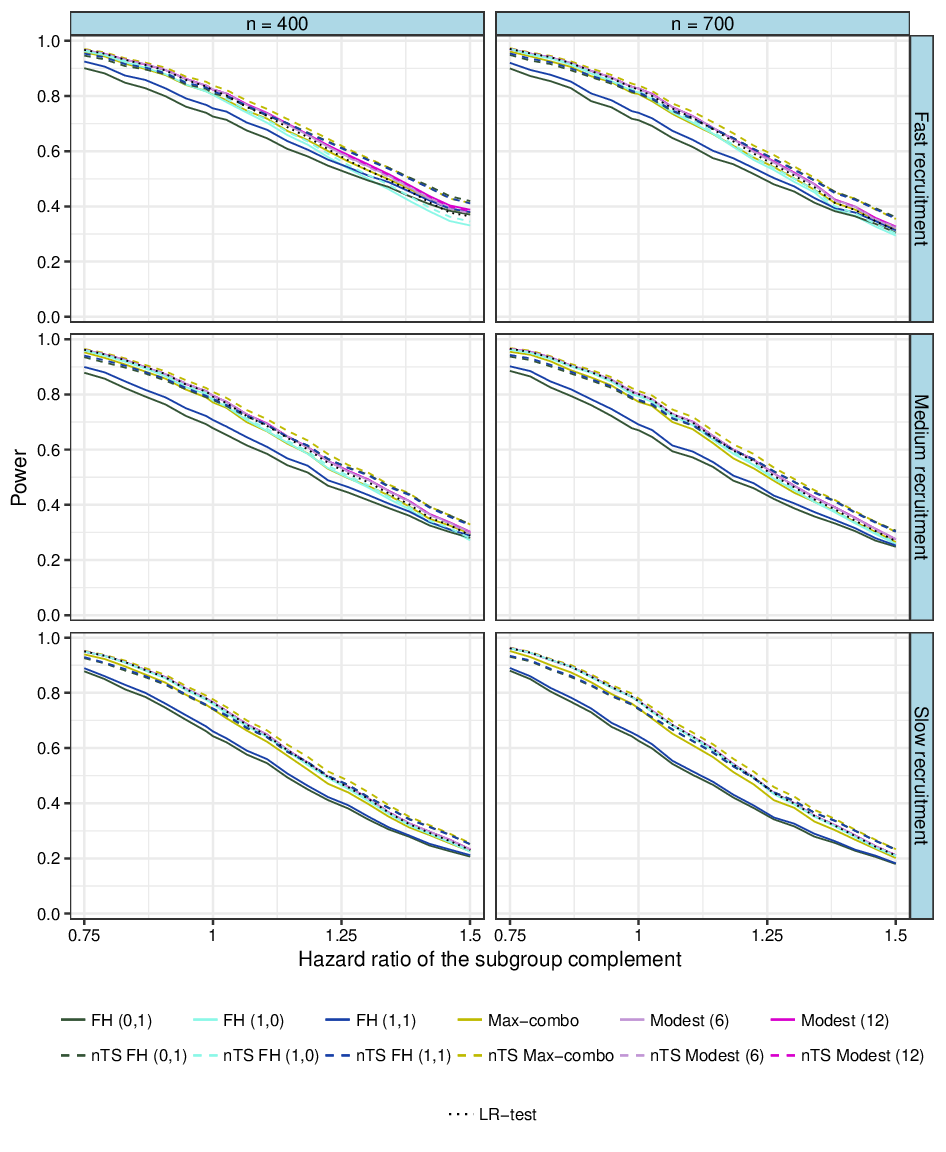}
 \caption{\textbf{High subgroup prevalence scenario:} Power of the naive two-step ("nTS") tests with a pre-test significance level of  $\alpha_{pre} = 0.2$. Results of the two-step tests are displayed as dashed lines in the same colour as the respective unconditional approaches under NPH. The power of the log-rank test ("LR-test") serves as a reference and is displayed as a black dotted line. The hazard ratio is with regard to the complement of the treatment subgroup. The HR of the treatment subgroup compared to the control is always 0.1 (corresponding to a median survival time of 120 months).}
	\label{res:sub50_alpha0.2}
\end{figure}

\begin{figure}[H]
	\centering
  \includegraphics[width=.87\textwidth]{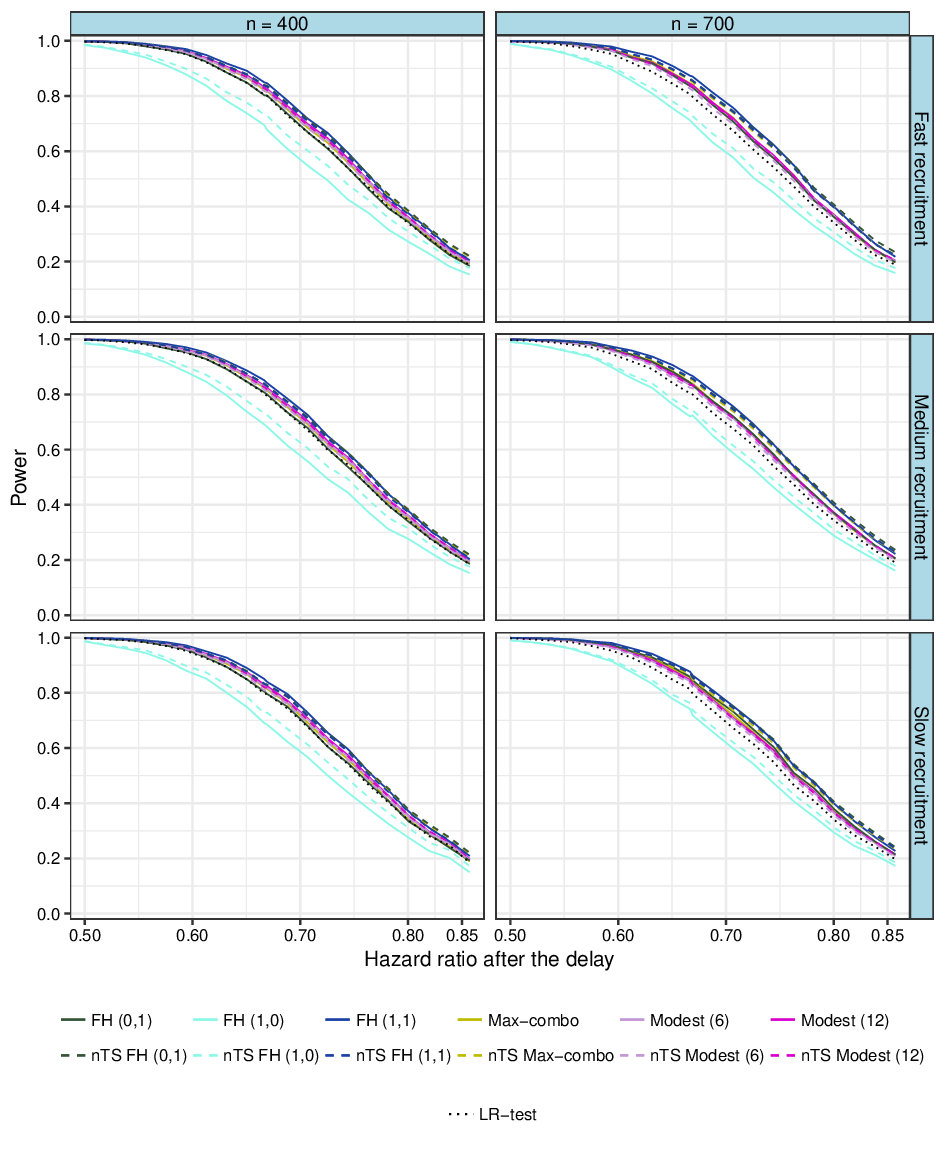}
 \caption{\textbf{Short delayed effect scenario:} Power of the naive two-step ("nTS") tests with a pre-test significance level of  $\alpha_{pre} = 0.2$. Results of the two-step tests are displayed as dashed lines in the same colour as the respective unconditional approaches under NPH. The power of the log-rank test ("LR-test") serves as a reference and is displayed as a black dotted line.}
	\label{res:delay2_alpha0.2}
\end{figure}

\begin{figure}[H]
	\centering
  \includegraphics[width=.87\textwidth]{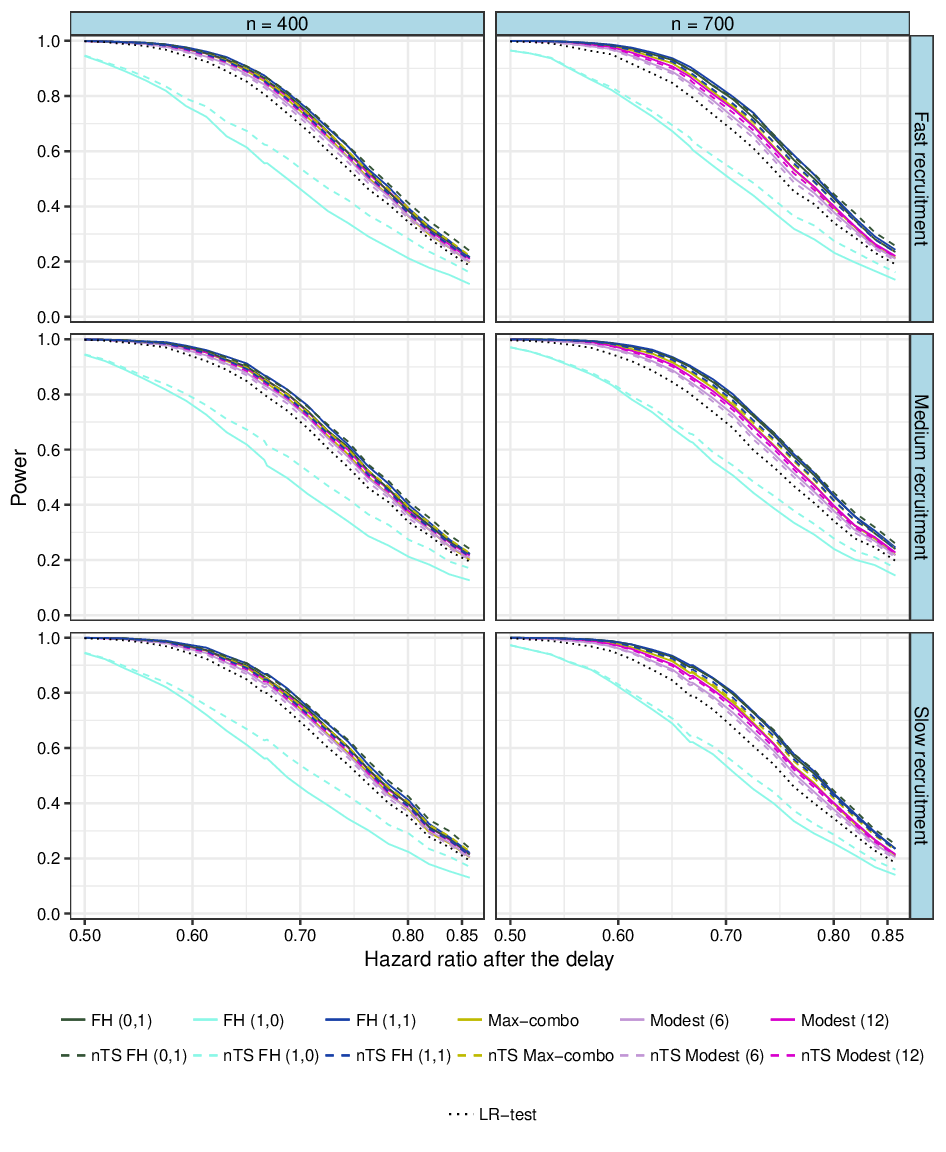}
 \caption{\textbf{Long delayed effect scenario:} Power of the naive two-step ("nTS") tests with a pre-test significance level of  $\alpha_{pre} = 0.2$. Results of the two-step tests are displayed as dashed lines in the same colour as the respective unconditional approaches under NPH. The power of the log-rank test ("LR-test") serves as a reference and is displayed as a black dotted line.}
	\label{res:delay4_alpha0.2}
\end{figure}

\begin{figure}[H]
	\centering
  \includegraphics[width=.87\textwidth]{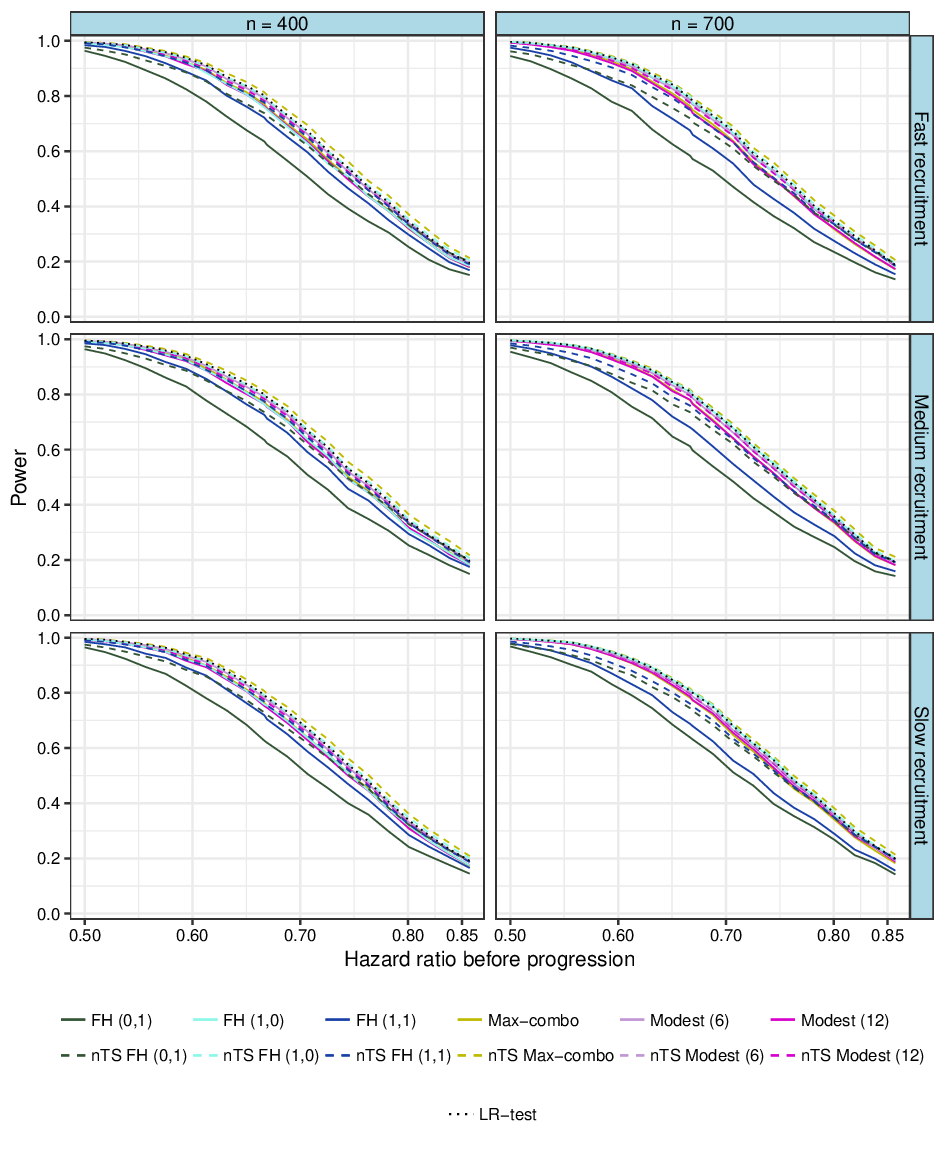}
 \caption{\textbf{Disease progression scenario (Short survival):} Power of the naive two-step ("nTS") tests with a pre-test significance level of  $\alpha_{pre} = 0.2$. Results of the two-step tests are displayed as dashed lines in the same colour as the respective unconditional approaches under NPH. The power of the log-rank test ("LR-test") serves as a reference and is displayed as a black dotted line.}
	\label{res:prog3_alpha0.2}
\end{figure}

\begin{figure}[H]
	\centering
  \includegraphics[width=.87\textwidth]{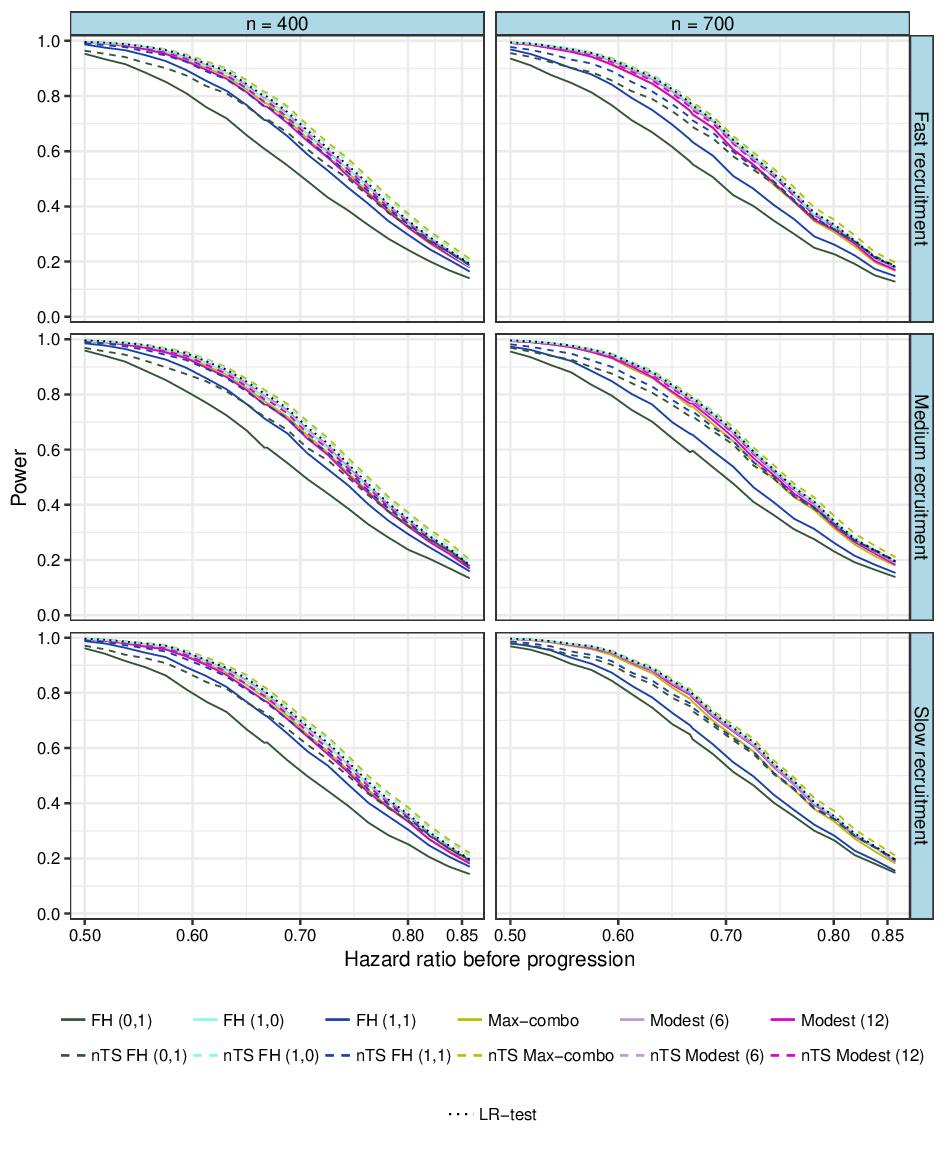}
 \caption{\textbf{Disease progression scenario (Long survival):} Power of the naive two-step ("nTS") tests with a pre-test significance level of  $\alpha_{pre} = 0.2$. Results of the two-step tests are displayed as dashed lines in the same colour as the respective unconditional approaches under NPH. The power of the log-rank test ("LR-test") serves as a reference and is displayed as a black dotted line.}
	\label{res:prog9_alpha0.2}
\end{figure}